\documentclass[12pt, draftclsnofoot, onecolumn]{IEEEtran}
\usepackage{amssymb}
\usepackage{amsmath}
\usepackage{cite}
\usepackage{url}
\usepackage{xcolor}
\usepackage{cite,graphicx,amsmath,amssymb}
\usepackage{subfigure}
\usepackage{citesort}
\usepackage{fancyhdr}
\usepackage{mdwmath}
\usepackage{mdwtab}
\usepackage{caption}
\usepackage{amsthm}
\usepackage{setspace}
\usepackage{algorithm}
\usepackage{algorithmic}
\usepackage{makecell}
\usepackage{diagbox}
\usepackage{multirow}

\newtheorem{remark}{Remark}
\newtheorem{theorem}{Theorem}

\newtheorem{lemma}{Lemma}

\newtheorem{corollary}{Corollary}

\newtheorem{proposition}{Proposition}
\newcommand{\bm}[1]{\mbox{\boldmath{$#1$}}}
\allowdisplaybreaks
\begin{document}

\title{Capacity and Optimal Resource Allocation for IRS-assisted Multi-user Communication Systems}
%

\author{

Xidong~Mu,~\IEEEmembership{Student Member,~IEEE,}
        Yuanwei~Liu,~\IEEEmembership{Senior Member,~IEEE,}
       Li~Guo,~\IEEEmembership{Member,~IEEE,}
       Jiaru~Lin,~\IEEEmembership{Member,~IEEE,}
       and Naofal~Al-Dhahir,~\IEEEmembership{Fellow,~IEEE}

\thanks{Part of this work will be presented at the IEEE International Conference on Communications (ICC), Montreal, Canada, June 14-23, 2021.~\cite{Mu2021}}
\thanks{X. Mu, L. Guo, and J. Lin are with the School of Artificial Intelligence and the Key Laboratory of Universal Wireless Communications, Ministry of Education, Beijing University of Posts and Telecommunications, Beijing, China. (email:\{muxidong, guoli, jrlin\}@bupt.edu.cn).}
\thanks{Y. Liu is with the School of Electronic Engineering and Computer Science, Queen Mary University of London, London, UK. (email:yuanwei.liu@qmul.ac.uk).}
\thanks{N. Al-Dhahir is with the Department of Electrical and Computer Engineering, The University of Texas at Dallas, Richardson, TX 75080 USA.(e-mail: aldhahir@utdallas.edu).}
}
\maketitle
\vspace{-1.8cm}
\begin{abstract}
\vspace{-0.5cm}
The fundamental capacity limits of intelligent reflecting surface (IRS)-assisted multi-user wireless communication systems are investigated in this paper. Specifically, the capacity and rate regions for both capacity-achieving non-orthogonal multiple access (NOMA) and orthogonal multiple access (OMA) transmission schemes are characterized by jointly optimizing the IRS reflection matrix and wireless resource allocation under the constraints of a maximum number of IRS reconfiguration times. In NOMA, all users are served in the same resource blocks by employing superposition coding and successive interference cancelation techniques. In OMA, all users are served by being allocated orthogonal resource blocks of different sizes. \textcolor{black}{For NOMA, the ideal case with an asymptotically large number of IRS reconfiguration times is firstly considered, where the optimal solution is obtained by employing the Lagrange duality method. Inspired by this result, an inner bound of the capacity region for the general case with a finite number of IRS reconfiguration times is derived. For OMA, the optimal transmission strategy for the ideal case is to serve each individual user alternatingly with its effective channel power gain maximized. Based on this result, a rate region inner bound for the general case is derived.} Finally, numerical results are provided to show that: i) a significant capacity and rate region improvement can be achieved by using IRS; ii) the capacity gain can be further improved by dynamically configuring the IRS reflection matrix.
\end{abstract}
\vspace{0.3cm}
\section{Introduction}
The rapid development of various advanced applications (e.g., extended reality, autonomous driving, etc.) imposes more requirements on the fifth-generation (5G) and beyond (B5G) wireless networks, including higher data rate, lower latency and higher reliability~\cite{Saad6G}. To meet those requirements, a variety of wireless technologies have been proposed, such as massive multiple-input multiple-output (MIMO) and millimeter-wave (mmWave) communications \cite{Andrews5G}. Despite achieving significant performance gains, these technologies also require higher hardware cost and energy consumption. To this end, intelligent reflecting surface (IRS) is emerging as a promising cost-effective and green solution \cite{WuTowards,LiangLISA,Basar,Renzo}.\\
\indent IRS (also referred to as reconfigurable intelligent surface (RIS)) technology has drawn tremendous attention from both academia and industry. An IRS is a planar array, which consists of a large number of passive reflecting elements. Each element can passively reflect the incident electromagnetic wave while changing its amplitude and phase shift~\cite{WuTowards,LiangLISA}. With an IRS smart controller, the reflected signal propagation can be artificially changed to enhance the network performance. For instance, if the transmitter and receiver are blocked by an obstacle, an extra path can be created with the deployment of the low-cost IRS. Due to the nearly passive full-duplex mode of operation, the IRS does not suffer the self-interference problem as compared with conventional relaying technologies such as amplify-and-forward (AF) and decode-and-forward (DF) relays~\cite{WuTowards}.
\vspace{-0.6cm}
\subsection{Prior Works}
\vspace{-0.3cm}
Growing research efforts have been devoted to investigate the performance gain of IRS under different objectives and application scenarios. For example, Wu {\em et al.}~\cite{Wu2019IRS} minimized the total transmit power by alternatively optimizing the active beamforming at the access point (AP) and the passive beamforming at the IRS. An IRS power consumption model was proposed by Huang {\em et al.}~\cite{Huang2019}, where the energy efficiency (EE) was maximized for an IRS-assisted downlink multi-user network. The achievable spectral efficiency was maximized by Yu {\em et al.}~\cite{Yu} in a single-user IRS-assisted multiple-input single-output (MISO) communication system, where the passive beamforming was designed using fixed point iteration and manifold optimization techniques. Yang {\em et al.}~\cite{Yang_OFDMA} proposed a dynamic passive beamforming scheme to maximize the minimum rate in an IRS-enhanced orthogonal frequency division multiple access (OFDMA) network. The channel capacity of an IRS-assisted MIMO system was maximized by Zhang {\em et al.}~\cite{Zhang_Capacity}, where alternating optimization algorithms were proposed under frequency-flat and frequency-selective channels. Guo {\em et al.}~\cite{Guo} investigated the weighted sum rate maximization problem under imperfect channel state information (CSI), where the active and passive beamforming were optimized by applying the stochastic successive convex approximation (SCA) algorithm. With the aim of achieving secrecy transmission, Chen {\em et al.}~\cite{Chen2019} proposed to deploy the IRS in a downlink MISO system coexisting with multiple eavesdroppers, where the passive beamforming was designed under different practical IRS elements constraints. Yu {\em et al.}~\cite{Yu_Robust} investigated IRS-assisted secure communications with imperfect CSI. Furthermore, the application of IRS in simultaneous wireless information and power transfer (SWIPT) systems was studied in \cite{Wu_SWIPT}, which revealed that dedicated energy signals are not required in the IRS-assisted SWIPT. Li {\em et al.}~\cite{IRS_UAV} studied the joint trajectory and passive beamforming optimization in IRS-assisted unmanned aerial vehicle (UAV) communications. The IRS effectiveness was evaluated in~\cite{Dai_ax} via experimental tests at 2.3 GHz and 28.5 GHz. \\
\indent To further improve the system performance, some initial studies have focused on the integration of IRS and NOMA technologies. Ding {\em et al.}~\cite{Ding} proposed to deploy IRSs to enhance the received signal strength of cell-edge users in NOMA transmission. Under this setup, the outage performance was analyzed under an on-off IRS control scheme. The max-min rate problem in the IRS-NOMA network was investigated by Yang {\em et al.}~\cite{Yang_ax}. Fu {\em et al.}~\cite{Fu} minimized the transmit power in a downlink IRS-assisted MISO system, where an efficient difference-of-convex (DC) programming based algorithm was proposed for passive beamforming designs. The sum rate of all users in an IRS-NOMA network was maximized in~\cite{Mu_ax} with ideal and non-ideal IRS element assumptions. Moreover, Zhu {\em et al.}~\cite{Zhu_ax} designed the passive beamforming with the concept of quasi-degradation condition and proposed a hybrid NOMA transmission scheme. Hou {\em et al.}~\cite{Hou_ax2} analyzed SE and EE performance of the IRS-assisted NOMA network with a priority based design. A theoretical performance comparison between NOMA and OMA was performed in~\cite{Zheng}, which showed that asymmetric and symmetric user pairing schemes are favored by NOMA and OMA, respectively.
\subsection{Motivations and Contributions}
\vspace{-0.3cm}
Multiple-access (MA) techniques are essential for integrating IRS into multi-user wireless communications. Although prior research contributions have considered frequency division multiple access (FDMA), time division multiple access (TDMA), and NOMA transmission schemes \cite{Yang_OFDMA,Yang_ax,Fu,Mu_ax,Zhu_ax,Zheng}, the obtained solutions were in general suboptimal from an information-theoretic perspective. Note that there is one prior work~\cite{Zhang_Capacity} that studied the capacity limits of IRS-assisted point-to-point MIMO systems. However, the results in~\cite{Zhang_Capacity} did not consider MA techniques and cannot be applied in the multi-user scenario. To the best of our knowledge, the fundamental capacity limits of IRS-assisted multi-user wireless communications and globally optimal transmission strategies have not been investigated yet. However, investigating these problems is of vital importance to determine system performance upper bounds and provide useful guidelines for practical system design, which motivates the main study of this work.\\
\indent Besides achieving a higher capacity, the combination of IRS and NOMA is also conceived to be a win-win strategy for wireless networks due to the following reasons:
\begin{itemize}
  \item \textbf{IRS to NOMA:} In conventional NOMA transmission, the SIC decoding orders among users are in general determined by their channel conditions which can not be modified artificially. With the help of IRSs, by properly adjusting the reflection coefficients, the reflected signals can be combined coherently or destructively with the non-reflected signal to enhance or degrade the effective channel power gains of users. As a result, NOMA decoding orders can be designed more freely. This unique degree-of-freedom (DoF) provided by IRSs enables a \emph{flexible} NOMA operation to be carried out, thus improving the performance of NOMA communication.
  \item \textbf{NOMA to IRS:} Facing the stringent communication requirements of future wireless networks and the problem of spectrum shortage, more flexible and efficient resource allocations can be facilitated by NOMA for IRS-assisted communications compared to conventional OMA. Thus, diversified communication requirements can be satisfied and the spectral efficiency can be further improved.
\end{itemize}

\indent Against above backgrounds, in this paper, we investigate IRS-assisted multi-user communication systems where a single-antenna AP sends independent information to multiple single-antenna users with the aid of one IRS. For practical implementation, the IRS uses discrete phase shifts. Different from the existing works assuming that the IRS reflection matrix is fixed through the entire transmission, in our work, it can be reconfigured $N$ times depending on the time duration for configuring the IRS. Under this setup, we jointly optimize the IRS reflection matrix as well as resource allocation to reveal the fundamental capacity limits of IRS-assisted multi-user wireless communications. The main contributions of this paper are as follows:
\begin{itemize}
  \item We characterize the capacity and rate regions for both capacity-achieving NOMA and OMA schemes. By utilizing the rate-profile technique, the Pareto boundary of these regions can be characterized by maximizing the average sum rate of all users, subject to a set of rate-profile constraints, discrete IRS phase shifts, the maximum number of IRS reconfiguration times, and resource allocation constraints.
  \item For the capacity region of NOMA, we first consider the ideal case with an asymptotically large number of IRS reconfiguration times, i.e., $N \to \infty $. \textcolor{black}{The formulated Pareto boundary characterization problem is shown to satisfy the time-sharing condition \cite{timeshare}, and thus can be globally optimally solved using the Lagrange duality method.} The derived optimal solution reveals that the optimal transmission strategy for NOMA is carrying out alternating transmission among different user groups and decoding orders. \textcolor{black}{Inspired by the obtained optimal solutions, we develop an efficient iterative algorithm to find the inner bound of the capacity region for the general case of finite $N$.}
  \item \textcolor{black}{For the rate region of OMA, we globally optimally solve the Pareto boundary characterization problem for $N \to \infty $.} The optimal transmission strategy for OMA is alternating transmission among each individual user with its corresponding effective channel power gain maximized. \textcolor{black}{Based on this result, we further find the inner bound of the rate region for finite $N$.}
  \item Our numerical results demonstrate that 1) both the capacity and rate regions achieved by introducing the IRS are significantly larger than those without the IRS; 2) dynamically reconfiguring the IRS reflection matrix can increase the capacity gain, especially for OMA; 3) the performance gain of NOMA over OMA in the IRS-assisted system outperforms than that without the IRS.
\end{itemize}
\vspace{-0.6cm}
\subsection{Organization and Notations}
\vspace{-0.3cm}
The rest of this paper is organized as follows. Section II presents the system model of the IRS-assisted multi-user communication system and the two transmission schemes, namely NOMA and OMA. Then, we characterize the Pareto boundary of the capacity and rate regions for NOMA and OMA in Section III and Section IV, respectively. Section V presents numerical results to demonstrate the performance of our proposed designs and compare them with other benchmark schemes. Finally, Section VI concludes the paper. \\
\indent \emph{Notations:} Scalars are denoted by lower-case letters. Vectors and matrices are denoted by bold-face lower-case and upper-case letters, respectively. ${\mathbb{C}^{N \times 1}}$ denotes the space of $N \times 1$ complex-valued vectors. ${{\mathbf{a}}^T}$, ${{\mathbf{a}}^H}$ and ${\rm {diag}}\left( \mathbf{a} \right)$ denote the transpose, the conjugate transpose and the diagonal matrix of vector ${\bf{a}}$, respectively. ${{\mathbf{0}}_N}$ denotes a $1 \times N$ vector whose elements are zero.
\vspace{-0.6cm}
\section{System Model and Transmission Schemes}
\vspace{-0.3cm}
\subsection{System Model}
\begin{figure}[h!]
    \begin{center}
        \includegraphics[width=2.5in]{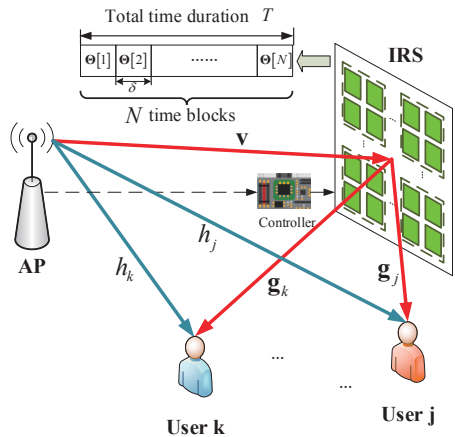}
        \caption{Illustration of the IRS-assisted multi-user communication system.}
        \label{System model}
    \end{center}
\end{figure}
\vspace{-1cm}
As shown in Fig. \ref{System model}, we consider an IRS-assisted multi-user communication system, where a single-antenna AP transmits independent information to $K$ single-antenna users with the aid of an IRS equipped with $M_R$ passive reflecting elements. The IRS is controlled by the AP through a smart controller. Since the IRS usually has a large number of passive reflecting elements, configuring the IRS can be highly complex and time-consuming. To address this issue, the adjacent IRS elements with high channel correlation are grouped into a sub-surface and share a common reflection coefficient, as assumed in~\cite{Zheng,Yang_CSI}. Let $B$ denote the size of each sub-surface. The IRS with $M_R$ passive reflecting elements is further divided into $M = \frac{{{M_R}}}{B}$ sub-surfaces. Fig. \ref{System model} illustrates the grouping scheme with $B=4$. In this paper, we assume that all channels follow the quasi-static block fading channel model \cite{Tse,Yang_OFDMA}, where the channel condition remains approximately constant in each channel coherence block. To reveal the most essential design insights and for ease of exposition, we focus on one specific channel coherence block and let $T$ denote the block duration. Furthermore, let $\delta $ denote the time duration required by the AP to configure the IRS and the total time duration $T$ can be further divided into $N = \left[ {\frac{T}{\delta }} \right]$ time blocks\footnote{In this paper, we assume that the users are static or moving slowly, which is also one of the most typical scenarios for the application of IRS. In this case, the channel coherence time $T$ is on the order of 25 ms~\cite{Tse}. In addition, as reported in~\cite{RFfocus}, the time duration $\delta $ is 0.22 ms - 7 ms depending on the number of IRS elements. Therefore, it is practical to assume that the IRS reflection matrix can be reconfigured multiple times. This new degree-of-freedom (DoF) has been initially investigated in some recent research contributions \cite{Yang_OFDMA,Yang_CSI,Karasik}.}. As a result, as illustrated in Fig. \ref{System model}, the IRS reflection matrix can be reconfigured only at the beginning of each time block $n \in {{\mathcal{N}}} = \left\{ {1,2, \ldots ,N} \right\}$ and remains fixed within each time block. It is worth mentioning that if $N=1$, the IRS reflection matrix is fixed through the whole transmission as assumed in the prior IRS research contributions.\\
\indent To characterize the capacity region with the IRS, we assume that the CSI of all channels involved can be perfectly obtained at the AP\footnote{The results with perfect CSI in this work actually provide a theoretical performance upper bound for the considered system.} with the recently proposed channel estimation methods~\cite{Yang_CSI,Nadeem_CSI}.  Let ${\mathbf{v}} \in {\mathbb{C}^{M \times 1}}$ and ${h_k}$ denote the corresponding AP-IRS channel and that between the AP and user $k$\footnote{Due to the ``double fading'' effect~\cite{Ozdogan}, the powers of the signals reflected by the IRS two or more times are much smaller than those of signals reflected one time, and thus can be ignored in this paper.}. In addition, the channel between the IRS and user $k$ is denoted by ${{\mathbf{g}}_k} \in {\mathbb{C}^{M \times 1}}$. At the $n$th time block, the IRS's diagonal reflection matrix is denoted by ${\mathbf{\Theta}} \left[ n \right] = {\rm{diag}}\left( {{\beta _1}\left[ n \right]{e^{j{\theta _1}\left[ n \right]}},{\beta _2}\left[ n \right]{e^{j{\theta _2}\left[ n \right]}}, \cdots ,{\beta _M}\left[ n \right]{e^{j{\theta _M}\left[ n \right]}}} \right)$, where ${{\beta _m}\left[ n \right]}$ and ${\theta _m}\left[ n \right] \in \left[ {0,2\pi } \right)$ are the amplitude and phase shift coefficients of the $m$th sub-surface, respectively. For practical implementation, we assume a finite resolution phase shift for each IRS element, which has a constant reflection amplitude (i.e., ${\beta _m}\left[ n \right] = 1,\forall n,m$) and discrete phase values ${{\mathcal{D}}} \triangleq \left\{ {\frac{{n2\pi }}{{{L}}},n = 0,1,2, \cdots ,{L} - 1} \right\}$, where $L=2^b$ and $b$ denotes the number of bits to adjust the phase. Let ${\mathcal{S}}$ denote the set of all possible phase-shift matrices at the IRS and $\left| {\mathcal{S}} \right| \triangleq {L^{M}}$.\\
\indent The combined channel power gain from the AP to user $k$ during the $n$th time block is given by ${\left| {{h_k} + {\mathbf{g}}_k^H{\mathbf{\Theta}} \left[ n \right]{\mathbf{v}}} \right|^2}$. Let $s_k\left[ n \right]$ and ${p _k}\left[ n \right]$ denote the transmitted information-bearing signal and the transmit power for user $k$ during the $n$th time block, respectively. Therefore, the received signal of user $k$ at the $n$th time block can be expressed as
\begin{align}\label{yk}
{y_k}\left[ n \right] = \left( {{h_k} + {\mathbf{g}}_k^H{\mathbf{\Theta}} \left[ n \right]{\mathbf{v}}} \right) \sum\nolimits_{k = 1}^K {\sqrt {{p_k}\left[ n \right]} {s_k}\left[ n \right]}  + {n_k}\left[ n \right],
\end{align}
where ${n_k}\left[ n \right]$ is the additive white Gaussian noise (AWGN) at user $k$. For ease of exposition, the noise power of each user is assumed to be equal to ${{\sigma ^2}}$ and the instantaneous power constraint at the AP is considered. Let ${P_{\max }}$ denote the maximum transmit power constraint, then we have $\sum\nolimits_{k = 1}^K {{p_k}\left[ n \right]}  \le {P_{\max }},\forall n$. With the aim of achieving the capacity region of this channel, the AP should employ Gaussian signaling by setting ${{s_k}\left[ n \right]}$'s as independent circularly symmetric complex Gaussian (CSCG) random variables with zero mean and unit variances ${\mathbb{E}}\left( {{{\left| {{s_k}\left[ n \right]} \right|}^2}} \right) = 1,\forall k$.
\vspace{-0.6cm}
\subsection{Capacity-achieving NOMA Transmission Scheme}
\vspace{-0.3cm}
\indent First, we consider the capacity-achieving NOMA transmission scheme~\cite{Ergodic}, where users share the same time and frequency resources by invoking superposition coding at the AP and successive interference cancelation (SIC) at the users~\cite{Liu2017,Ding2017}. Based on the NOMA principle, each user employs SIC to remove the co-channel interference. The user with a stronger channel power gain can decode the signal of the user with weaker channel power gain. Let ${\mu _k}\left[ n \right]$ denote the decoding order for user $k$ at time block $n$. For instance, if ${\mu _k}\left[ n \right] = i$, then user $k$ is the $i$th signal to be decoded. For any two users $j$ and $k$ satisfying ${\mu _j} \left[ n \right] < {\mu _k} \left[ n \right]$, the combined channel power gains of the two users need to satisfy the condition that ${\left| {{h_k} + {\mathbf{g}}_k^H{\mathbf{\Theta}} \left[ n \right]{\mathbf{v}}} \right|^2} \ge {\left| {{h_j} + {\mathbf{g}}_j^H{\mathbf{\Theta}} \left[ n \right]{\mathbf{v}}} \right|^2}$. With this condition, it can be verified that the decoding rate at user $k$ to decode the signal of user $j$ is always no less than the data rate at user $j$ to decode its own signal, and thus SIC can be successfully performed for the given decoding order~\cite{Liu2018Multiple}. Therefore, the achievable rate in bits per second per Hertz (bit/s/Hz) of user $k$ at the $n$th time block in the NOMA scheme is given by
\vspace{-0.4cm}
\begin{align}\label{rate NOMA}
R_k^{{\rm{N}}}\left[ n \right] = {\log _2}\left( {1 + \frac{{{{\left| {{h_k} + {\mathbf{g}}_k^H{\mathbf{\Theta}} \left[ n \right]{\mathbf{v}}} \right|}^2}{p_k}\left[ n \right]}}{{\sum\nolimits_{{\mu _i}\left[ n \right] > {\mu _k}\left[ n \right]} {{{\left| {{h_k} + {\mathbf{g}}_k^H{\mathbf{\Theta}} \left[ n \right]{\mathbf{v}}} \right|}^2}{p_i}\left[ n \right]}  + {\sigma ^2}}}} \right).
\end{align}
\vspace{-1cm}

\noindent Then, the average achievable rate of user $k$ over the entire period $T$ in the NOMA scheme is $\overline R_k^{{\rm{N}}} = \frac{1}{N}\sum\nolimits_{n = 1}^{{{N}}} {R_k^{{\rm{N}}}\left[ n \right]}$.
\vspace{-0.6cm}
\subsection{OMA Transmission Scheme}
\vspace{-0.3cm}
For the OMA transmission scheme, e.g., frequency division multiple access (FDMA) or time division multiple access (TDMA), the $k$th user receives its information ${s_k}\left[ n \right]$ with the transmit power ${p_k}\left[ n \right]$ over ${\omega _k}\left[ n \right] \in \left[ {0,1} \right]$ of the total orthogonal resources (time/frequency) at the $n$th time block, where $\sum\nolimits_{k = 1}^K {{\omega _k}} \left[ n \right] \le 1,\forall n$. As mentioned before, the IRS reflection matrix ${\mathbf{\Theta}}\left[ n \right]$ can be reconfigured only at the beginning of each time block. All users for both FDMA and TDMA share the identical ${\mathbf{\Theta}}\left[ n \right]$ for each time block. Then, the achievable rate of user $k$ at the $n$th time block in the OMA scheme can be expressed as
\vspace{-0.3cm}
\begin{align}\label{rate OMA}
R_k^{\rm{O}}\left[ n \right] = {\omega _k}\left[ n \right]{\log _2}\left( {1 + \frac{{{{\left| {{h_k} + {\mathbf{g}}_k^H{\mathbf{\Theta}}\left[ n \right]{\mathbf{v}}} \right|}^2}{p_k}\left[ n \right]}}{{{\omega _k}\left[ n \right]{\sigma ^2}}}} \right)
\end{align}
\vspace{-0.8cm}

\noindent Note that the expression in \eqref{rate OMA} is applicable to both FDMA and TDMA scenarios since the consumed energy in TDMA at each time block (given by $ {\sum\nolimits_{k = 1}^K {{\omega _k}\left[ n \right]\frac{{{p_k}\left[ n \right]}}{{{\omega _k}\left[ n \right]}}} } $) is the same as that in FDMA (given by $ {\sum\nolimits_{k = 1}^K {{p_k}\left[ n \right]} } $). Similarly, the average achievable rate of user $k$ over the entire period $T$ in the OMA scheme is given by $\overline R_k^{\rm{O}} = \frac{1}{N}\sum\nolimits_{n = 1}^{{{N}}} {R_k^{\rm{O}}\left[ n \right]} $.
\vspace{-0.6cm}
\section{Capacity Region Characterization with NOMA}
\vspace{-0.3cm}
In this section, we investigate the capacity region\footnote{As the NOMA transmission scheme has been shown to be capacity-achieving in \cite{Tse}, in this paper, we define the capacity region to be the set of average achievable rate-tuples over the considered channel coherence duration $T$, which can be simultaneously achievable by all users for NOMA. A similar definition is also applied for the rate region with OMA in Section IV.} for the NOMA transmission scheme. Let ${{\mathcal{X}}^{{\rm{N}}}}$ denote the feasible sets of $\left\{ {{\mathbf{\Theta}} \left[ n \right],{p_k}\left[ n \right],\forall n} \right\}$ specified by the discrete phase shift values and the maximum total transmit power constraint. Accordingly, the capacity region achieved by NOMA is defined
as\cite{Ergodic}
\vspace{-0.4cm}
\begin{align}\label{rate region NOMA}
{{\mathcal{C}}}\left( {b,N} \right) \triangleq \mathop  \cup \limits_{\left\{ {{\mathbf{\Theta}} \left[ n \right],{p_k}\left[ n \right]} \right\} \in {{{\mathcal{X}}}^{{\rm{N}}}}} \overline {{\mathcal{C}}} \left( {\left\{ {{\mathbf{\Theta}} \left[ n \right],{p_k}\left[ n \right]} \right\}} \right),
\end{align}
\vspace{-1cm}

\noindent where ${\overline {{\mathcal{C}}}}\left( {\left\{ {{\mathbf{\Theta}} \left[ n \right],{p_k}\left[ n \right]} \right\}} \right) = \left\{ {\overline {\mathbf{r}}:0 \le {\overline r_k} \le \overline R_k^{{\rm{N}}},\forall k} \right\}$ denotes the set of all achievable average rate-tuples $\overline {\mathbf{r}} \triangleq \left( {{\overline r_1},{\overline r_2}, \cdots ,{\overline r_K}} \right)$ for all $K$ users under given $\left\{ {{\mathbf{\Theta}} \left[ n \right],{p_k}\left[ n \right],\forall n} \right\}$.\\
\indent From the definition in \eqref{rate region NOMA}, ${{\mathcal{C}}}\left( {b,N} \right)$ consists of the set of average rate-tuples for all users that can be simultaneously achieved over the period $T$ with the NOMA transmission scheme. The upper-right boundary of this region is called the \emph{Pareto boundary}, at which it is impossible to improve the rate of one user without simultaneously decreasing the rate of the other users. In order to characterize the complete Pareto boundary, we invoke the \emph{rate-profile} technique \cite{rate_profile}, which is guaranteed to find all Pareto boundary points even if the region is a non-convex set. Specifically, let ${\bm \alpha}  = \left[ {{\alpha _1},{\alpha _2}, \cdots ,{\alpha _K}} \right]$ denote a rate-profile vector, where ${{\alpha _k}}$ represents the rate allocation among the $K$ users. We have $\sum\nolimits_{k = 1}^K {{\alpha _k}}  = 1$ and ${\alpha _k} \ge 0,\forall k$. Then, the characterization of any Pareto boundary point of the capacity region ${{\mathcal{C}}}\left( {b,N} \right)$ is formulated as the following optimization problem
\vspace{-0.6cm}
\begin{subequations}
\begin{align}\label{P1 NOMA}
({{\rm{P1}}}):&\mathop {\max }\limits_{R^{{\rm{N}}},\overline {\mathbf{r}},\left\{ {{\mathbf{\Theta}} \left[ n \right],{p_k}\left[ n \right]} \right\}} \;\;R^{{\rm{N}}}  \\
\label{rate allocation NOMA}{\rm{s.t.}}\;\;&{\overline r_k} \ge {\alpha _k}R^{{\rm{N}}},\forall k,\\
\label{rate2 NOMA}&\overline {\mathbf{r}} \in {\overline {{\mathcal{C}}}^{{\rm{N}}}}\left( {\left\{ {{\mathbf{\Theta}} \left[ n \right],{p_k}\left[ n \right]} \right\}} \right),\\
\label{discrete phase shift NOMA}&{\mathbf{\Theta}}\left[ n \right] \in {\mathcal{S}},\forall n,\\
\label{total power1 NOMA}&\sum\nolimits_{k = 1}^K { {{p_k}\left[ n \right]} } \le {P_{\max}},\forall n,\\
\label{total power2 NOMA}&{p_k}\left[ n \right] \ge 0,\forall k,n,\\
\label{SIC condition}&
  {\left| {{h_k} + {\mathbf{g}}_k^H{\mathbf{\Theta}} \left[ n \right]{\mathbf{v}}} \right|^2} \ge {\left| {{h_j} + {\mathbf{g}}_j^H{\mathbf{\Theta}} \left[ n \right]{\mathbf{v}}} \right|^2},{\rm{if}}\;{\mu _j}\left[ n \right] < {\mu _k}\left[ n \right]\forall k,j,n,
\end{align}
\end{subequations}
\vspace{-1.2cm}

\noindent where ${R^{{\rm{N}}}}$ denotes the average achievable sum rate of the $K$ users in the NOMA transmission scheme. Constraints \eqref{discrete phase shift NOMA} and \eqref{total power1 NOMA} are the the discrete phase-shift matrix constraint and total transmit power constraint, respectively. \eqref{SIC condition} denotes the user decoding order constraint.\\
\indent \textcolor{black}{It is worth noting that due to the average sum rate objective function, ${R^{\rm{N}}}$, and the rate profile constraints of, problem (P1) can not be directly decomposed into $N$ independent subproblems, each of them represents one specific time block. Without lose of optimality, the optimization variables over each time block should be jointly optimized for maximizing the average sum rate subject to the rate profile constraints.} Moreover, problems (P1) is a highly-coupled non-convex problem due to the non-convex set ${\mathcal{S}}$, and the non-convex constraints \eqref{rate2 NOMA} and \eqref{SIC condition}. To solve this problem, we first characterize the capacity region by considering the total number of time blocks is asymptotically large, i.e., $N \to \infty $. Then, we investigate the capacity region inner bound with any finite value $N$.
\vspace{-0.7cm}
\subsection{Capacity Region: $N \to \infty $}
\vspace{-0.3cm}
\indent In this subsection, we investigate problem (P1) when $N \to \infty $, where the corresponding capacity region is denoted by ${{\mathcal{C}}}\left( {b,\infty } \right)$. This can be regarded as an ideal case, where the time duration for configuring the IRS reflection matrix is negligible, i.e., $\delta  \to 0$. Before solving problem (P1), we first have the following theorem.
\vspace{-0.4cm}
\begin{theorem}\label{tsproof}
\emph{Problem (P1) satisfies the time-sharing condition \cite{timeshare} when $N \to \infty $.}
\begin{proof}
See Appendix A.
\end{proof}
\end{theorem}
\vspace{-0.5cm}

\noindent \textbf{Theorem 1} shows that problem (P1) satisfies the time-sharing condition when $N \to \infty $. According to the convex analysis in \cite{timeshare}, in this case, the strong duality \cite{convex} holds and the duality gap between the primal problem and its Lagrange dual problem is zero. Hence, we can derive the optimal solution to (P1) via its dual problem.\\
\indent Next, we invoke the Lagrange duality method to optimally solve (P1) with asymptotically large $N$. By utilizing the Lagrange duality method, the partial Lagrangian function of problem (P1) can be expressed as
\vspace{-0.5cm}
\begin{align}\label{Lagrange D_NOMA}
\begin{gathered}
  {{{\mathcal{L}}}_1}\left( {{R_\infty ^{{\rm{N}}}},{\mathbf{\Theta}} \left[ n \right],\left\{ {{p_k}\left[ n \right]} \right\},\left\{ {\lambda _k^{{\rm{N}}}} \right\}} \right) = \left( {1 - \sum\nolimits_{k = 1}^K {{\alpha _k}\lambda _k^{{\rm{N}}}} } \right){R_\infty ^{{\rm{N}}}} \hfill \\
   + \sum\nolimits_{k = 1}^K {\frac{{\lambda _k^{{\rm{N}}}}}{N}\sum\nolimits_{n = 1}^{{{N}}} {{{\log }_2}\left( {1 + \frac{{{{\left| {{h_k} + {\mathbf{g}}_k^H{\mathbf{\Theta}} \left[ n \right]{\mathbf{v}}} \right|}^2}{p_k}\left[ n \right]}}{{\sum\nolimits_{{\mu _i}\left[ n \right] > {\mu _k}\left[ n \right]} {{{\left| {{h_k} + {\mathbf{g}}_k^H{\mathbf{\Theta}} \left[ n \right]{\mathbf{v}}} \right|}^2}{p_i}\left[ n \right]}  + {\sigma ^2}}}} \right)} } , \hfill \\
\end{gathered}
\end{align}
\vspace{-0.8cm}

\noindent where ${\left\{ {{\lambda _k^{{\rm{N}}}}} \right\}}$ are the non-negative Lagrange multipliers associated with constraint \eqref{rate allocation NOMA}. Accordingly, the Lagrange dual function of problem (P1) is given by
\vspace{-0.4cm}
\begin{subequations}\label{Lagrange dual function D_NOMA}
\begin{align}
&{f_1}\left( {\left\{ {\lambda _k^{{\rm{N}}}} \right\}} \right) = \mathop {\max }\limits_{{R_\infty ^{{\rm{N}}}},{\mathbf{\Theta}} \left[ n \right],\left\{ {{p_k}\left[ n \right]} \right\}} {{{\mathcal{L}}}_1}\left( {R_\infty^{{\rm{N}}},{\mathbf{\Theta}} \left[ n \right],\left\{ {{p_k}\left[ n \right]} \right\},\left\{ {\lambda _k^{{\rm{N}}}} \right\}} \right)  \\
\label{rate allocation D_NOMA}&{\rm{s.t.}}\;\;\eqref{discrete phase shift NOMA}-\eqref{SIC condition}.
\end{align}
\end{subequations}
\vspace{-1.6cm}
\begin{lemma}\label{zero}
\emph{In order for the dual function ${f_1}\left( {\left\{ {\lambda _k^{{\rm{N}}}} \right\}} \right)$ to be upper-bounded from above, i.e., ${f_1}\left( {\left\{ {\lambda _k^{{\rm{N}}}} \right\}} \right) < + \infty $, it must hold that $\sum\nolimits_{k = 1}^K {{\alpha _k}{\lambda _k} }  = 1$.}
\begin{proof}
\emph{This is shown by contradiction. Suppose that $\sum\nolimits_{k = 1}^K {{\alpha _k}{\lambda _k^{{\rm{N}}}}}  > 1$ or $\sum\nolimits_{k = 1}^K {{\alpha _k}{\lambda _k^{{\rm{N}}}}}  < 1$. Then, by setting ${R_\infty ^{{\rm{N}}}} \to  - \infty $ or ${R_\infty ^{{\rm{N}}}} \to  + \infty $, we have ${f_1}\left( {\left\{ {\lambda _k^{{\rm{N}}}} \right\}} \right) \to  + \infty $. Therefore, neither of the above two inequalities can be true and the lemma is proved.}
\end{proof}
\end{lemma}
\vspace{-0.4cm}
Based on \textbf{lemma \ref{zero}}, the dual problem of problem (P1) is given by
\vspace{-0.4cm}
\begin{subequations}\label{dual problem D_NOMA}
\begin{align}
\left({{\rm{D1}}}\right):&\mathop {\min }\limits_{\left\{ {\lambda _k^{{\rm{N}}}} \right\}} \;\;\;{f_1}\left( {\left\{ {\lambda _k^{{\rm{N}}}} \right\}} \right) \\
\label{dual1 D_NOMA}{\rm{s.t.}}\;\;&{\sum\nolimits_{k = 1}^K {{\alpha _k}\lambda _k^{{\rm{N}}}}  = 1},{\lambda _k^{{\rm{N}}} \ge 0},\forall k.
\end{align}
\end{subequations}
\vspace{-1.2cm}

\noindent As the strong duality holds, we can optimally solve problem (P1) by solving its dual problem (D1). In the following, we first solve problem \eqref{Lagrange dual function D_NOMA} to obtain $f_1\left( {\left\{ {{\lambda _k^{{\rm{N}}}}} \right\} } \right)$ under any given dual variables, then solve problem (D1) to find the optimal dual variables ${\left\{ {{\lambda _k^{*{\rm{N}}}}} \right\}  }$ to minimize $f_1\left( {\left\{ {{\lambda _k^{{\rm{N}}}}} \right\} } \right)$, and finally construct the optimal primal solution to problem (P1).
\subsubsection{Obtaining $f_1\left( {\left\{ {{\lambda _k^{{\rm{N}}}}} \right\}} \right)$ by Solving Problem \eqref{Lagrange dual function D_NOMA}}
In order to obtain ${f_1}\left( {\left\{ {\lambda _k^{{\rm{N}}}} \right\}} \right)$ for given dual variables ${\left\{ {\lambda _k^{{\rm{N}}}} \right\}}$, we set $R_\infty^{*{\rm{N}}} = 0$ and drop the time block index $n$. Then, problem \eqref{Lagrange dual function D_NOMA} can be expressed as
\vspace{-0.4cm}
\begin{subequations}\label{D NOMA p}
\begin{align}\label{D NOMA}
\mathop {\max }\limits_{{\mathbf{\Theta}},\left\{ {{p_k}} \right\} } &\sum\nolimits_{k = 1}^K {\frac{{\lambda _k^{{\rm{N}}}}}{T}} {\log _2}\left( {1 + \frac{{{{\left| {{h_k} + {\mathbf{g}}_k^H{\mathbf{\Theta}} {\mathbf{v}}} \right|}^2}{p_k}}}{{\sum\nolimits_{{\mu _i} > {\mu _k}} {{{\left| {{h_k} + {\mathbf{g}}_k^H{\mathbf{\Theta}} {\mathbf{v}}} \right|}^2}{p_i}}  + {\sigma ^2}}}} \right)  \\
\label{con D NOMA3}{\rm{s.t.}}\;\;&{{\mathbf{\Theta}}} \in {\mathcal{S}},\\
\label{con D NOMA1}&\sum\nolimits_{k = 1}^K { {{p_k}} } \le {P_{\max}},\\
\label{con D NOMA20}&{p_k} \ge 0,\forall k,\\
\label{con D NOMA2}&{\left| {{h_k} + {\mathbf{g}}_k^H{\mathbf{\Theta}} {\mathbf{v}}} \right|^2} \ge {\left| {{h_j} + {\mathbf{g}}_j^H{\mathbf{\Theta}} {\mathbf{v}}} \right|^2},{\rm {if}}\;{\mu _j} < {\mu _k}.
\end{align}
\end{subequations}
\vspace{-1.2cm}

\noindent Problem \eqref{D NOMA p} can be regarded as a weighted sum rate maximization problem. The optimal solution is achieved when \eqref{con D NOMA1} is satisfied with equality, since otherwise we can always increase the power allocation to the strongest user $p_K$ to increase the cost function. For ease of exposition, we assume that the decoding order is ${\mu _k} \triangleq k,\forall k$ and define ${q_k} = \sum\nolimits_{i = k}^K {{p_i}} ,\forall k$, where ${q_1}=P_{\max}$. The $k$th term in \eqref{D NOMA} can be expressed as
\vspace{-0.4cm}
\begin{align}\label{qk}
  \begin{gathered}
  {\log _2}\left( {1 + \frac{{{{\left| {{h_k} + {\mathbf{g}}_k^H{\mathbf{\Theta}} {\mathbf{v}}} \right|}^2}{p_k}}}{{\sum\limits_{i > k} {{{\left| {{h_k} + {\mathbf{g}}_k^H{\mathbf{\Theta}} {\mathbf{v}}} \right|}^2}{p_i}}  + {\sigma ^2}}}} \right) \hfill \\
  \;\;\;\;\;\;\;\;\;\; = {\log _2}\left( {{\sigma ^2} + {{\left| {{h_k} + {\mathbf{g}}_k^H{\mathbf{\Theta}} {\mathbf{v}}} \right|}^2}{q_k}} \right) - {\log _2}\left( {{\sigma ^2} + {{\left| {{h_k} + {\mathbf{g}}_k^H{\mathbf{\Theta}} {\mathbf{v}}} \right|}^2}{q_{k + 1}}} \right). \hfill \\
\end{gathered}
\end{align}
\vspace{-0.8cm}

\indent Next, we first focus on the weighted sum rate maximization problem under any given dual variables ${\left\{ {\lambda _k^{{\rm{N}}}} \right\}}$ and IRS reflection matrix ${\mathbf{\Theta}}$. Let ${\phi ^{\left( {\left\{ {\lambda _k^{{\rm{N}}}} \right\},{\mathbf{\Theta}} } \right)}}\left( {\left\{ {{q_k}} \right\}} \right)$ denote the corresponding objective function, the optimization problem can be expressed as
\vspace{-0.3cm}
\begin{subequations}\label{D NOMA qk}
\begin{align}
\mathop {\max }\limits_{\left\{ {{q_k}} \right\}} \;\;&{\phi ^{\left( {\left\{ {\lambda _k^{{\rm{N}}}} \right\},{\mathbf{\Theta}} } \right)}}\left( {\left\{ {{q_k}} \right\}} \right)  \\
\label{con D NOMA qk1}{\rm{s.t.}}\;\;&{P_{\max }} = {q_1} \ge {q_2} \ge  \cdots  \ge {q_K} \ge 0,
\end{align}
\end{subequations}
\vspace{-1.2cm}

\noindent where ${\phi ^{\left( {\left\{ {\lambda _k^{{\rm{N}}}} \right\},{\mathbf{\Theta}} } \right)}}\left( {\left\{ {{q_k}} \right\}} \right)$ is expressed as
\vspace{-0.3cm}
\begin{align}\label{phiphi}
\begin{gathered}
  {\phi ^{\left( {\left\{ {\lambda _k^{{\rm{N}}}} \right\},{\mathbf{\Theta}} } \right)}}\left( {\left\{ {{q_k}} \right\}} \right) = \frac{{\lambda _1^{{\rm{N}}}}}{T}{\log _2}\left( {{\sigma ^2} + {{\left| {{h_1} + {\mathbf{g}}_1^H{\mathbf{\Theta}} {\mathbf{v}}} \right|}^2}{q_1}} \right) - \frac{{\lambda _K^{{\rm{N}}}}}{T}{\log _2}\left( {{\sigma ^2}} \right) \hfill \\
  + \sum\nolimits_{k = 2}^K {\left( {\frac{{\lambda _k^{{\rm{N}}}}}{T}{{\log }_2}\left( {{\sigma ^2} + {{\left| {{h_k} + {\mathbf{g}}_k^H{\mathbf{\Theta}} {\mathbf{v}}} \right|}^2}{q_k}} \right) - \frac{{\lambda _{k - 1}^{{\rm{N}}}}}{T}{{\log }_2}\left( {{\sigma ^2} + {{\left| {{h_{k - 1}} + {\mathbf{g}}_{k - 1}^H{\mathbf{\Theta}} {\mathbf{v}}} \right|}^2}{q_k}} \right)} \right)}.  \hfill \\
\end{gathered}
\end{align}
\vspace{-0.4cm}

\noindent Since ${\phi ^{\left( {\left\{ {\lambda _k^{{\rm{N}}}} \right\},{\mathbf{\Theta}} } \right)}}\left( {\left\{ {{q_k}} \right\}} \right)$ is a continuous function over the feasible region $\Psi  = \left\{ {{q_k},\forall k|{P_{\max}} = {q_1}} \right.$\\$\left. { \ge {q_2} \ge  \cdots  \ge {q_K} \ge 0} \right\}$, its maximum point is either at the stationary point or on the boundary of $\Psi$. To solve problem \eqref{D NOMA qk}, we have the following lemma.
\vspace{-0.4cm}
\begin{lemma}\label{2K-1}
\emph{For the $K$ user case, the number of candidate solutions for achieving the maximum of ${\phi ^{\left( {\left\{ {\lambda _k^{{\rm{N}}}} \right\},{\mathbf{\Theta}} } \right)}}\left( {\left\{ {{q_k}} \right\}} \right)$ is ${2^K} - 1$.}
\begin{proof}
\emph{The inequality sign in the constraint \eqref{con D NOMA qk1} can be further decomposed into equality and strict inequality. As a result, the original constraint ${P_{\max }} = {q_1}\underbrace  \ge _1{q_2}\underbrace  \ge _2 \cdots \underbrace  \ge _{K - 1}{q_K}\underbrace  \ge _K0$ can be replaced with ${2^{K - 1}}$ independent constraints since it is infeasible for the case ${P_{\max }} = {q_1} = {q_2} =  \cdots  = {q_K} = 0$. Therefore, there are ${2^K} - 1$ candidate solutions associated with each decomposed constraint to achieve the maximum of ${\phi ^{\left( {\left\{ {\lambda _k^{{\rm{N}}}} \right\},{\mathbf{\Theta}} } \right)}}\left( {\left\{ {{q_k}} \right\}} \right)$. The proof is completed.}
\end{proof}
\end{lemma}
\vspace{-0.6cm}
\textbf{Lemma \ref{2K-1}} provides important insights on how to maximize ${\phi ^{\left( {\left\{ {\lambda _k^{{\rm{N}}}} \right\},{\mathbf{\Theta}} } \right)}}\left( {\left\{ {{q_k}} \right\}} \right)$ based on the constraint \eqref{con D NOMA qk1}. From the definition of $q_k$, if ${q_k} = {q_{k + 1}}$, it follows that the $k$th user is not served (i.e., ${p_k} = 0$); otherwise the $k$th user is served with ${p_k} > 0$. On this basis, we derive the optimal solution of the two user and three user cases using the following proposition.
\vspace{-0.4cm}
\begin{proposition}\label{two user NOMA}
\emph{The optimal power allocation to problem \eqref{D NOMA qk} with two users is given by}
\vspace{-0.3cm}
\begin{align}\label{q2}
  \left( {q_1^*,q_2^*} \right) = \arg\max \left\{ {{\phi ^{\left( {\left\{ {\lambda _k^{{\rm{N}}}} \right\},{\mathbf{\Theta}} } \right)}}\left( {{P_{\max }},0} \right),} \right. \left. {{\phi ^{\left( {\left\{ {\lambda _k^{{\rm{N}}}} \right\},{\mathbf{\Theta}} } \right)}}\left( {{P_{\max }},{P_{\max }}} \right),{\phi ^{\left( {\left\{ {\lambda _k^{{\rm{N}}}} \right\},{\mathbf{\Theta}} } \right)}}\left( {{P_{\max }},{{\overline q }_2}} \right)} \right\}
\end{align}
\vspace{-1cm}

\noindent \emph{and with three users is given by}
\vspace{-0.2cm}
\begin{align}\label{q3}
\begin{gathered}
  \left( {q_1^*,q_2^*,q_3^*} \right) = \arg\max \left\{ {{\phi ^{\left( {\left\{ {\lambda _k^{{\rm{N}}}} \right\},{\mathbf{\Theta}} } \right)}}\left( {{P_{\max }},0,0} \right),}{\phi ^{\left( {\left\{ {\lambda _k^{{\rm{N}}}} \right\},{\mathbf{\Theta}} } \right)}}\left( {{P_{\max }},{P_{\max }},0} \right), \right. \hfill \\
  {\phi ^{\left( {\left\{ {\lambda _k^{{\rm{N}}}} \right\},{\mathbf{\Theta}} } \right)}}\left( {{P_{\max }},{P_{\max }},{P_{\max }}} \right),{\phi ^{\left( {\left\{ {\lambda _k^{{\rm{N}}}} \right\},{\mathbf{\Theta}} } \right)}}\left( {{P_{\max }},{P_{\max }},{{\overline q }_3}} \right),{\phi ^{\left( {\left\{ {\lambda _k^{{\rm{N}}}} \right\},{\mathbf{\Theta}} } \right)}}\left( {{P_{\max }},{{\overline q }_2},0} \right), \hfill \\
  \left. {{\phi ^{\left( {\left\{ {\lambda _k^{{\rm{N}}}} \right\},{\mathbf{\Theta}} } \right)}}\left( {{P_{\max }},{{\overline q }_3},{{\overline q }_3}} \right),{\phi ^{\left( {\left\{ {\lambda _k^{{\rm{N}}}} \right\},{\mathbf{\Theta}} } \right)}}\left( {{P_{\max }},{{\overline q }_2},{{\overline q }_3}} \right)} \right\}, \hfill \\
\end{gathered}
\end{align}
\vspace{-0.8cm}

\noindent \emph{where}
\begin{align}\label{q0}
{\overline q _k} = \left[ {\frac{{\left( {\frac{{\lambda _{k - 1}^{{\rm{N}}}}}{{{{\left| {{h_k} + {\mathbf{g}}_k^H{\mathbf{\Theta}} {\mathbf{v}}} \right|}^2}}} - \frac{{\lambda _k^{{\rm{N}}}}}{{{{\left| {{h_{k - 1}} + {\mathbf{g}}_{k - 1}^H{\mathbf{\Theta}} {\mathbf{v}}} \right|}^2}}}} \right)}}{{\lambda _k^{{\rm{N}}} - \lambda _{k - 1}^{{\rm{N}}}}}} \right]_0^{{P_{\max }}}, \forall k.
\end{align}
\begin{proof}
See Appendix B.
\end{proof}
\end{proposition}
\vspace{-0.4cm}
Therefore, we can optimally solve problem \eqref{D NOMA qk} by checking all candidate solutions. Based on problem \eqref{D NOMA qk}, we adopt exhaustive search over the IRS reflection matrix set ${\mathcal{S}}$ to obtain the optimal IRS reflection matrix to problem \eqref{D NOMA p} under given ${\left\{ {\lambda _k^{{\rm{N}}}} \right\}}$ as
\vspace{-0.4cm}
\begin{align}\label{theta NOMA}
{{\mathbf{\Theta}} ^*} = \arg \mathop {\max }\limits_{{\mathbf{\Theta}}  \in {{\mathcal{S}}}} \left\{ {\mathop {\max }\limits_{\left\{ {{q_k}} \right\}} \;\;{\phi ^{\left( {\left\{ {\lambda _k^{{\rm{N}}}} \right\},{\mathbf{\Theta}} } \right)}}\left( {\left\{ {{q_k}} \right\}} \right)} \right\}.
\end{align}
\vspace{-1cm}

\noindent Accordingly, the optimal power allocation solutions to problem \eqref{D NOMA p} under given ${\left\{ {\lambda _k^{{\rm{N}}}} \right\}}$ are given by
\vspace{-0.6cm}
\begin{align}\label{pk NOMA}
\begin{gathered}
  {p_k^{*}} = q_k^{*\left( {\left\{ {\lambda _k^{{\rm{N}}}} \right\},{\mathbf{\Theta}}^* } \right)} - q_{k + 1}^{*\left( {\left\{ {\lambda _k^{{\rm{N}}}} \right\},{\mathbf{\Theta}}^* } \right)},1 \le k \le K - 1, \hfill \\
  {p_K^*} = q_K^{*\left( {\left\{ {\lambda _k^{{\rm{N}}}} \right\},{\mathbf{\Theta}}^* } \right)}. \hfill \\
\end{gathered}
\end{align}
\vspace{-0.8cm}

\noindent By substituting the above optimal solutions $\left\{ {{\mathbf{\Theta}}^* ,\left\{ {p_k^*} \right\}} \right\}$ into problem \eqref{Lagrange dual function D_NOMA}, the dual function ${f_1}\left( {\left\{ {\lambda _k^{{\rm{N}}}} \right\}} \right)$ is obtained.
\subsubsection{Finding Optimal Dual Solution to (D1)} Next, we search over ${\left\{ {{\lambda _k^{{\rm{N}}}}} \right\}}$ to minimize $f_1\left( {\left\{ {{\lambda _k^{{\rm{N}}}}} \right\}} \right)$ for solving (D1). Since the dual problem (D1) is always convex but in general non-differentiable, the subgradient-based methods such as the ellipsoid method \cite{Ellipsoid} can be used to solve problem (D1). Note that the subgradient of the objective function $f_1\left( {\left\{ {{\lambda _k^{{\rm{N}}}}} \right\}} \right)$ is denoted by ${{\mathbf{s}}_0} = { {\Delta {\bm{{\lambda }}}} }$, where $\Delta{\lambda _k} = {\log _2}\left( {1 + \frac{{{{\left| {{h_k} + {\mathbf{g}}_k^H{{\mathbf{\Theta}} ^*}{\mathbf{v}}} \right|}^2}p_k^*}}{{\sum\nolimits_{{\mu _i} > {\mu _k}} {{{\left| {{h_k} + {\mathbf{g}}_k^H{{\mathbf{\Theta}} ^*}{\mathbf{v}}} \right|}^2}p_i^*}  + {\sigma ^2}}}} \right),\forall k$. Moreover, the equality constraint \eqref{dual1 D_NOMA} is equivalent to the two inequality constraints: $1 - \sum\nolimits_{k = 1}^K {{\alpha _k}{\lambda _k} }  \le 0$ and $ - 1 + \sum\nolimits_{k = 1}^K {{\alpha _k}{\lambda _k} }  \le 0$, whose subgradients are given by ${{\mathbf{s}}_1} = -{\bm \alpha}$ and ${{\mathbf{s}}_2} = -{{\mathbf{s}}_1}$. With the above subgradients, the dual variables can be updated by the constrained ellipsoid method. The optimal dual solutions to (D1) are denoted by ${\left\{ {\lambda _k^{*{\rm{N}}}} \right\}}$.
\subsubsection{Constructing Optimal Primal Solution to Problem (P1)} With the obtained optimal dual variable ${\left\{ {\lambda _k^{*{\rm{N}}}} \right\}}$ using the constrained ellipsoid method, we need to find the optimal primal solutions to problem (P1). It is worth noting that when using the Lagrange dual method to solve a convex problem via its dual problem, the optimal solution which maximizes the Lagrange function under the optimal dual solution is the optimal primal solution if and only if such a solution is unique and primal feasible \cite{convex}. In our case, the optimal solutions ${\mathbf{\Theta}}^*, \left\{ {p_k^*} \right\}$ and $R_\infty^{*{{\rm{N}}}}$ to problem \eqref{Lagrange dual function D_NOMA} with ${\left\{ {\lambda _k^{*{\rm{N}}}} \right\}}$ are generally non-unique, additional steps are required to construct the optimal primal solution by deciding the time-sharing ratio among all optimal solutions. Suppose that problem \eqref{Lagrange dual function D_NOMA} under ${\left\{ {\lambda _k^{*{\rm{N}}}} \right\}}$ has a total number of $\Pi $ optimal solutions, denoted by $\left\{ {\left( {{{\mathbf{\Theta}}_\varpi ^{*} },\left\{ {p_{k,\varpi}^{*} } \right\}} \right)} \right\}_{\varpi  = 1}^\Pi $. Let ${\tau _\varpi }$ denote the optimal transmission duration at the $\varpi$th optimal solution. Then, the optimal primal solution to (P1) with asymptotically large $N$ can be obtained by solving the following problem
\vspace{-0.5cm}
\begin{subequations}\label{P1-NOMA}
\begin{align}
&\mathop {\max }\limits_{{R_\infty ^{{\rm{N}}}},\left\{ {{\tau _\varpi} \ge 0} \right\}} \;\;{R_\infty ^{{\rm{N}}}} \\
\label{P1-NOMA1}{\rm{s.t.}}\;\;&
  \sum\nolimits_{\varpi  = 1}^\Pi  {\frac{{{\tau _\varpi }}}{T}} {\log _2}\left( {1 + \frac{{{{\left| {{h_k} + {\mathbf{g}}_k^H{{\mathbf{\Theta}} _\varpi^* }{\mathbf{v}}} \right|}^2}p_{k,\varpi}^* }}{{\sum\nolimits_{\mu _i^\varpi  > \mu _k^\varpi } {{{\left| {{h_k} + {\mathbf{g}}_k^H{{\mathbf{\Theta}} _\varpi^* }{\mathbf{v}}} \right|}^2}p_{i,\varpi}^* }  + {\sigma ^2}}}} \right) \ge {\alpha _k}{R_\infty ^{{\rm{N}}}},\forall k, \\
\label{P1-NOMA2}&\sum\nolimits_{\varpi = 1}^\Pi {{\tau _\varpi}}  = T.
\end{align}
\end{subequations}
\vspace{-1.2cm}

\noindent Similarly, problem \eqref{P1-NOMA} is a standard standard linear program (LP), which can be solved by using standard convex optimization tools such as CVX \cite{cvx}. As a result, the optimal solution to problem (P1) is obtained. The details of the procedures for optimally solving problem (P1) are summarized in \textbf{Algorithm 1}. The computational complexity of Algorithm 1 is dominated by the ellipsoid method in steps 1)-4) and solving the LP problem \eqref{P1-NOMA}. Specifically, the complexity of steps 2)-3) is ${\mathcal{O}}\left( {{L^M}{K^2}} \right)$. As the ellipsoid method requires ${\mathcal{O}}\left( {\frac{{{K^2}}}{\varepsilon }} \right)$ to converge \cite{Ellipsoid}, the total complexity for steps 1)-4) is ${\mathcal{O}}\left( {\frac{{{L^M}{K^4}}}{\varepsilon }} \right)$. The complexity of solving problem \eqref{P1-NOMA} is ${\mathcal{O}}\left( {{{\left| \Pi  \right|}^3}} \right)$. Therefore, the total complexity for optimally solving (P1) is ${\mathcal{O}}\left( {\frac{{{L^M}{K^4}}}{\varepsilon } + {{\left| \Pi  \right|}^3}} \right)$.
\begin{algorithm}[!t]\label{method1}
\caption{Algorithm for Optimally Solving Problem (P1) when $N \to \infty $}
 \hspace*{0.02in}
\hspace*{0.02in} {Initialize an ellipsoid ${\mathcal{E}}\left( {{\left\{ {\lambda _k^{\rm{N}}} \right\} } ,{\mathbf{A}}} \right)$ containing $ {\left\{ {\lambda _k^{*{\rm{N}}}} \right\}} $, where ${{\left\{ {\lambda _k^{{\rm{N}}}} \right\} } }$ is the center point of ${\mathcal{E}}$ and the positive definite matrix ${\mathbf{A}}$ characterizes the size of ${\mathcal{E}}$.}\\
\vspace{-0.4cm}
\begin{algorithmic}[1]
\STATE {\bf repeat}
\STATE Obtain $R_\infty^{*{\rm{N}}},{\mathbf{\Theta}} ^*,\left\{ {p_k^*} \right\}$ based on \eqref{theta NOMA} and \eqref{pk NOMA}.
\STATE Update ${ {\left\{ {\lambda _k^{{\rm{N}}}} \right\}} }$ using the constrained ellipsoid with the corresponding subgradients.
\STATE {\bf until} ${ {\left\{ {\lambda _k^{{\rm{N}}}} \right\}} }$ converge with a prescribed accuracy.
\STATE Set ${\left\{ {\lambda _k^{*{\rm{N}}}} \right\}}  \leftarrow  {\left\{ {\lambda _k^{{\rm{N}}}} \right\} } $.
\STATE Obtain $\left\{ {\left( {{{\mathbf{\Theta}}_\varpi ^* },\left\{ {p_{k,\varpi}^* } \right\}} \right)} \right\}_{\varpi  = 1}^\Pi $ by solving problem \eqref{D NOMA p} under ${\left\{ {\lambda _k^{*{\rm{N}}}} \right\}}$.
\STATE Construct the optimal solution $R_\infty^{*{\rm{N}}}$ to problem (P1) via time-sharing by solving problem \eqref{P1-NOMA}.
\end{algorithmic}
\end{algorithm}
\vspace{-0.4cm}
\begin{remark}\label{Alternative transmission NOMA}
\emph{The optimal solution to (P1) with asymptotically large $N$ means that to achieve any point on the Pareto boundary of ${{\mathcal{C}}}\left( {b,\infty } \right)$, the optimal strategy for the NOMA scheme is alternating transmission among different user groups or decoding orders with the designed IRS reflection matrix.}
\end{remark}
\vspace{-1cm}
\subsection{Capacity Region Inner Bound with Finite $N$}
\vspace{-0.3cm}
In this subsection, we consider the general Pareto boundary characterization problem (P1) with finite value $N$. In this case, the time-sharing condition does not hold. One solution for problem (P1) is to exhaustively search over all possible configurations of the IRS reflection matrix over different time slots, and then solve the remaining resource allocation problem. However, the computational complexity for checking all possible IRS reflection matrix configurations is ${\mathcal{O}}\left( {{L^{MN}}} \right)$, which is unacceptable even for moderate $M$ or $N$. To tackle this obstacle, we propose a suboptimal algorithm\footnote{The exhaustive search based method for solving (P1) with finite value $N$ is used as a baseline scheme in Section V to verify the optimality of the proposed suboptimal algorithm.} motivated by the optimal solution obtained previously for the ideal case. Therefore, an inner bound of the capacity region ${{\mathcal{C}}}\left( {b,N } \right)$ can be derived efficiently.
\subsubsection{IRS Reflection Matrix Design over Finite $N$ Time Blocks}
\begin{figure}[t!]
    \begin{center}
        \includegraphics[width=2.4in]{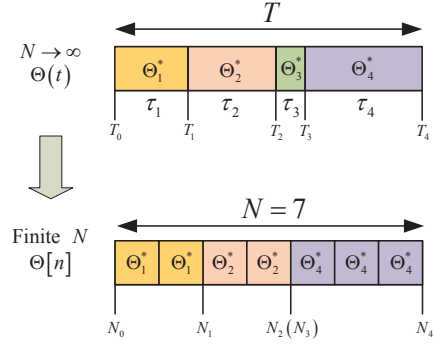}
        \caption{Illustration of the design of IRS reflection matrix for $N=7$ and $\Pi=4$.}
        \label{finite N}
    \end{center}
\end{figure}

Recall that the optimal solution to problem (P1) with asymptotically large $N$ corresponds to $\Pi $ optimal IRS reflection matrices and time duration $\left\{ {{\mathbf{\Theta}} _\varpi ^*,{\tau _\varpi }} \right\}_{\varpi  = 1}^\Pi $, which can be further expressed as
\vspace{-0.4cm}
\begin{align}\label{theta T}
\left\{ {{\mathbf{\Theta}} \left( t \right) = {\mathbf{\Theta}} _\varpi ^*,t \in \left[ {{T_{\varpi  - 1}},{T_\varpi }} \right)} \right\}_{\varpi  = 1}^\Pi,
\end{align}
\vspace{-1.2cm}

\noindent where ${T_\varpi } \triangleq \sum\nolimits_{i = 0}^\varpi  {{\tau _i}} $ and ${{\tau _0} \triangleq 0}$. Based on these, we construct the IRS reflection matrix over finite $N$ time blocks as follows
\vspace{-0.4cm}
\begin{align}\label{theta N}
\left\{ {{\mathbf{\Theta}} \left[ n \right] = {\mathbf{\Theta}} _\varpi ^*,{N_{\varpi  - 1}} + 1 \le n \le {N_\varpi }} \right\}_{\varpi  = 1}^\Pi,
\end{align}
\vspace{-1.2cm}

\noindent where ${N_\varpi } \triangleq \left[ {\frac{{{T_\varpi }}}{T}N} \right]$ and $\left[  \cdot  \right]$ denotes rounding to the nearest integer. It is worth noting that for any $\varpi $, if ${N_\varpi } < {N_{\varpi  - 1}} + 1$, then the IRS reflection coefficient ${{\mathbf{\Theta}} _\varpi ^*}$ is not adopted. Therefore, we obtain the IRS reflection matrix over finite $N$ time blocks, which is denoted by $\left\{ {{\mathbf{\Theta}}^* \left[ n \right]} \right\}_{n = 1}^{{{N}}}$. Fig. \ref{finite N} illustrates the design of the IRS reflection matrix with finite $N$.
\subsubsection{Capacity Region Inner Bound Characterization} With the constructed IRS reflection matrix $\left\{ {{\mathbf{\Theta}}^* \left[ n \right]} \right\}_{n = 1}^{{{N}}}$, problem (P1) can be expressed as the following power allocation problem
\vspace{-0.4cm}
\begin{subequations}\label{P1 sub}
\begin{align}
&\mathop {\max }\limits_{R^{{\rm{N}}},\left\{ {{p_k}\left[ n \right]} \right\}} \;\;R^{{\rm{N}}}  \\
\label{rate allocation NOMA sub}{\rm{s.t.}}\;\;&\frac{1}{N}\sum\nolimits_{n = 1}^{{{N}}} {R_k^{{\rm{N}}}\left[ n \right]}  \ge {\alpha _k}{R^{{\rm{N}}}},\\
\label{total power1 NOMA sub}&\sum\nolimits_{k = 1}^K { {{p_k}\left[ n \right]} } \le {P_{\max}},\forall n,\\
\label{total power2 NOMA sub}&{p_k}\left[ n \right] \ge 0,\forall k,n,
\end{align}
\end{subequations}
\vspace{-1cm}

\noindent where $R_k^{{\rm{N}}}\left[ n \right] = {\log _2}\left( {1 + \frac{{{H_k}\left[ n \right]{p_k}\left[ n \right]}}{{\sum\nolimits_{{\mu _i}\left[ n \right] > {\mu _k}\left[ n \right]} {{H_k}\left[ n \right]{p_i}\left[ n \right]}  + {\sigma ^2}}}} \right)$ and ${H_k}\left[ n \right] \triangleq {\left| {{h_k} + {\mathbf{g}}_k^H{\mathbf{\Theta}}^* \left[ n \right]{\mathbf{v}}} \right|^2},\forall k,n$. However, problem \eqref{P1 sub} is still non-convex due to the non-convex constraint \eqref{rate allocation NOMA sub}. To tackle it, $R_k^{{\rm{N}}}\left[ n \right]$ can be further expressed as
\vspace{-0.4cm}
\begin{align}\label{Rkn}
R_k^{{\rm{N}}}\left[ n \right] = {\log _2}\left( {{H_k}\left[ n \right]{P_{i,k}}\left[ n \right] + {\sigma ^2}} \right) - {\log _2}\left( {{H_k}\left[ n \right]{Q_{i,k}}\left[ n \right] + {\sigma ^2}} \right),
\end{align}
\vspace{-1.2cm}

\noindent where ${P_{i,k}}\left[ n \right] \triangleq \sum\nolimits_{{\mu _i}\left[ n \right] \ge {\mu _k}\left[ n \right]} {{p_i}\left[ n \right]} $ and ${Q_{i,k}}\left[ n \right] \triangleq \sum\nolimits_{{\mu _i}\left[ n \right] > {\mu _k}\left[ n \right]} {{p_i}\left[ n \right]} ,\forall i,k,n$. Note that $R_k^{{\rm{N}}}\left[ n \right]$ is the difference of two concave functions. By applying the first-order Taylor expansion, a concave lower bound at given local points $\left\{ {Q_{i,k}^{\left( l \right)}\left[ n \right]} \right\}$ can be expressed as
\vspace{-0.4cm}
\begin{align}\label{Rkn lb}
\begin{gathered}
  R_k^{{\rm{N}}}\left[ n \right] \ge R_{k,lb}^{{\rm{N}}}\left[ n \right] = {\log _2}\left( {{H_k}\left[ n \right]{P_{i,k}}\left[ n \right] + {\sigma ^2}} \right) \hfill \\
   - {\log _2}\left( {{H_k}\left[ n \right]Q_{i,k}^{\left( l \right)}\left[ n \right] + {\sigma ^2}} \right) - \frac{{{H_k}\left[ n \right]{{\log }_2}e}}{{{H_k}\left[ n \right]Q_{i,k}^{\left( l \right)}\left[ n \right] + {\sigma ^2}}}\left( {Q_{i,k}^{}\left[ n \right] - Q_{i,k}^{\left( l \right)}\left[ n \right]} \right). \hfill \\
\end{gathered}
\end{align}
\vspace{-0.8cm}

\noindent By replacing the non-convex terms in \eqref{rate allocation NOMA sub} with their concave lower bound, problem \eqref{P1 sub} can be written as
\vspace{-0.6cm}
\begin{subequations}\label{P1 sub ap}
\begin{align}
&\mathop {\max }\limits_{{R^{{\rm{N}}}},\left\{ {{P_{i,k}}\left[ n \right],Q_{i,k}\left[ n \right]} \right\}} \;\;{R^{{\rm{N}}}}  \\
\label{rate allocation NOMA sub ap}{\rm{s.t.}}\;\;&\frac{1}{N}\sum\nolimits_{n = 1}^{{{N}}} {R_{k,lb}^{{\rm{N}}}\left[ n \right]}  \ge {\alpha _k}{R^{{\rm{N}}}},\\
\label{total power1 NOMA sub ap}&\eqref{total power1 NOMA sub},\eqref{total power2 NOMA sub}.
\end{align}
\end{subequations}
\vspace{-1.2cm}

\noindent Now, it can be verified that problem \eqref{P1 sub ap} is a convex problem, which can be efficiently solved by using standard convex optimization tools such as CVX \cite{cvx}. It is worth noting that due to the adoption of the global lower bounds in \eqref{Rkn lb}, the obtained objective value in problem \eqref{P1 sub ap} in general serves as a lower bound for that in problem \eqref{P1 sub}. The solutions obtained in each iteration $l$ are used as the input local points for the next iteration $l+1$ and the objective function of problem \eqref{P1 sub} behaves in a non-decreasing manner. Since problem \eqref{P1 sub} has a finite optimal value, the proposed iterative algorithm is guaranteed to converge to a locally optimal solution of problem \eqref{P1 sub}. After convergence, a high-quality inner bound of the capacity region for finite $N$ can be efficiently obtained. The computational complexity for solving problem \eqref{P1 sub ap} is ${{\mathcal{O}}}\left( {I{{\left( {2{K^2}N + 1} \right)}^{3.5}}} \right)$~\cite{convex}, where ${2{K^2}N + 1}$ is the number of optimization variables of \eqref{P1 sub ap} and $I$ denotes the number of iterations needed for convergence. As the proposed suboptimal approach first constructs the IRS reflection matrix configuration based on the results from \textbf{Algorithm 1}, the total computational complexity is ${{\mathcal{O}}}\left( {\frac{{{L^M}{K^4}}}{\varepsilon } + {{\left| \Pi  \right|}^3} + I{{\left( {2{K^2}N + 1} \right)}^{3.5}}} \right)$.
\vspace{-0.6cm}
\section{Rate Region Characterization with OMA}
\vspace{-0.3cm}
In this section, we investigate the rate region with the OMA transmission scheme. Let ${{\mathcal{X}}^{{\rm{O}}}}$ denote the feasible set of $\left\{ {{\mathbf{\Theta}} \left[ n \right],{p_k}\left[ n \right],{\omega _k}\left[ n \right],\forall n} \right\}$ specified by the discrete phase shift values, the maximum total transmit power constraint and the total orthogonal resources constraint. Then, the achievable rate region for OMA is defined as\cite{Ergodic}
\vspace{-0.3cm}
\begin{align}\label{rate region OMA1}
{{{\mathcal{R}}}}\left( {b,N} \right) \triangleq \mathop  \cup \limits_{\left\{ {{\mathbf{\Theta}} \left[ n \right],{p_k}\left[ n \right],{\omega _k}\left[ n \right]} \right\} \in {{{\mathcal{X}}}^{\rm{O}}}} {\overline {{\mathcal{R}}}}\left( {\left\{ {{\mathbf{\Theta}} \left[ n \right],{p_k}\left[ n \right],{\omega _k}\left[ n \right]} \right\}} \right),
\end{align}
\vspace{-0.7cm}

\noindent where ${\overline {{\mathcal{R}}}}\left( {\left\{ {{\mathbf{\Theta}} \left[ n \right],{p_k}\left[ n \right],{\omega _k}\left[ n \right]} \right\}} \right) = \left\{ {\overline {\mathbf{r}}:0 \le {\overline r_k} \le \overline R_k^{\rm{O}},\forall k} \right\}$ denotes the set of all achievable average rate-tuples $\overline {\mathbf{r}} \triangleq \left( {{\overline r_1},{\overline r_2}, \cdots ,{\overline r_K}} \right)$ for all $K$ users under given $\left\{ {{\mathbf{\Theta}} \left[ n \right],{p_k}\left[ n \right],{\omega _k}\left[ n \right],\forall n} \right\}$. In order to characterize the Pareto boundary of the rate region ${{{\mathcal{R}}}}\left( {b,N} \right)$, we still invoke the rate-profile technique. Under the rate-profile vector ${\bm \alpha}  = \left[ {{\alpha _1},{\alpha _2}, \cdots ,{\alpha _K}} \right]$, the Pareto boundary point of the rate region ${{{\mathcal{R}}}}\left( {b,N} \right)$ can be characterized by solving the following problem
\vspace{-0.5cm}
\begin{subequations}
\begin{align}\label{P1 OMA}
({{\rm{P2}}}):&\mathop {\max }\limits_{R^{\rm{O}}, \overline {\mathbf{r}} ,\left\{ {{\mathbf{\Theta}} \left[ n \right],{p_k}\left[ n \right],{\omega _k}\left[ n \right]} \right\}} \;\;R^{\rm{O}}  \\
\label{rate allocation OMA}{\rm{s.t.}}\;\;&{\overline r_k} \ge {\alpha _k}R^{\rm{O}},\forall k,\\
\label{rate2 OMA}&\overline {\mathbf{r}} \in {\overline {{\mathcal{C}}} ^{\rm{O}}}{}\left( {\left\{ {{\mathbf{\Theta}} \left[ n \right],{p_k}\left[ n \right],{\omega _k}\left[ n \right]} \right\}} \right),\\
\label{discrete phase shift OMA}&{\mathbf{\Theta}}\left[ n \right] \in {\mathcal{S}},\forall n,\\
\label{total power1 OMA}&\sum\nolimits_{k = 1}^K { {{p_k}\left[ n \right]} } \le {P_{\max}},\forall n,\\
\label{resource allocation1 OMA}&\sum\nolimits_{k = 1}^K {{\omega _k}} \left[ n \right] \le 1,\forall n,\\
\label{resource allocation2 OMA}&{p_k}\left[ n \right] \ge 0, 0 \le {\omega _k}\left[ n \right] \le 1,\forall k,n,
\end{align}
\end{subequations}
\vspace{-1.3cm}

\noindent where ${R^{\rm{O}}}$ denotes the average achievable sum rate of the $K$ users in the OMA transmission scheme. Constraint \eqref{discrete phase shift OMA} represents the discrete phase-shift matrix constraint. Constraint \eqref{total power1 OMA} and constraints \eqref{resource allocation1 OMA} are the total transmit power and orthogonal resources constraints.\\
\indent Due to the non-convex set ${\mathcal{S}}$ and the non-convex constraint \eqref{rate2 OMA}, problems (P2) is still a non-convex problem. In the following, we solve problem (P2) in both asymptotically large $N$ and finite $N$ cases.
\vspace{-0.6cm}
\subsection{Rate Region: $N \to \infty $}
\vspace{-0.3cm}
In this subsection, we characterize the rate region with OMA when $N \to \infty $, where the corresponding rate region is denoted by
${{\mathcal{R}}}\left( {b,\infty } \right)$. Similar to \textbf{Theorem 1}, it can be shown that problem (P2) with asymptotically large $N$ also satisfies the time-sharing condition. We still derive the optimal solution via its dual problem.\\
\indent By utilizing the Lagrange duality method, the partial Lagrangian function of problem (P2) can be expressed as
\vspace{-0.4cm}
\begin{align}\label{Lagrangian D OMA}
\begin{gathered}
  {{{\mathcal{L}}}_2}\left( {R_\infty ^{\rm{O}},{\mathbf{\Theta}} \left[ n \right],\left\{ {{p_k}\left[ n \right],{\omega _k}\left[ n \right]} \right\},\left\{ {\lambda _k^{\rm{O}}} \right\}} \right) = \left( {1 - \sum\nolimits_{k = 1}^K {{\alpha _k}\lambda _k^{\rm{O}}} } \right)R_\infty ^{\rm{O}} \hfill \\
   + \sum\nolimits_{k = 1}^K {\frac{{\lambda _k^{\rm{O}}}}{N}\sum\nolimits_{n = 1}^{{{N}}} {{\omega _k}\left[ n \right]{{\log }_2}\left( {1 + \frac{{{{\left| {{h_k} + {\mathbf{g}}_k^H{\mathbf{\Theta}} \left[ n \right]{\mathbf{v}}} \right|}^2}{p_k}\left[ n \right]}}{{{\omega _k}\left[ n \right]{\sigma ^2}}}} \right)} }  \hfill \\
\end{gathered}
\end{align}
\vspace{-0.9cm}

\noindent where ${\left\{ {{\lambda _k^{{\rm{O}}}}} \right\}}$ are the non-negative Lagrange multipliers associated with constraint \eqref{rate allocation OMA}. Accordingly, the Lagrange dual function of problem (P2) is given by
\vspace{-0.4cm}
\begin{subequations}\label{Lagrange dual function D_OMA}
\begin{align}
&{f_2}\left( {\left\{ {\lambda _k^{{\rm{O}}}} \right\}} \right) = \mathop {\max }\limits_{R_\infty^{{\rm{O}}},{\mathbf{\Theta}} \left[ n \right],\left\{ {{p_k}\left[ n \right],{\omega _k}\left[ n \right]} \right\}} {{{\mathcal{L}}}_2}\left( {R_\infty^{{\rm{O}}},{\mathbf{\Theta}} \left[ n \right],\left\{ {{p_k}\left[ n \right],{\omega _k}\left[ n \right]} \right\},\left\{ {\lambda _k^{{\rm{O}}}} \right\}} \right)\\
\label{rate allocation D_OMA cons}&{\rm{s.t.}}\;\;\eqref{discrete phase shift OMA}-\eqref{resource allocation2 OMA}.
\end{align}
\end{subequations}
\vspace{-1.2cm}

\noindent Similarly, the condition that ${\sum\nolimits_{k = 1}^K {{\alpha _k}\lambda _k^{{\rm{O}}} = 1} }$ must be satisfied to ensure that ${f_2}\left( {\left\{ {\lambda _k^{{\rm{O}}}} \right\}} \right)$ is bounded from above. Then, the dual problem of problem (P2) is given by
\vspace{-0.5cm}
\begin{subequations}
\begin{align}\label{dual problem D_OMA}
\left({{\rm{D2}}}\right):&\mathop {\min }\limits_{\left\{ {\lambda _k^{{\rm{O}}}} \right\}} \;\;\;{f_2}\left( {\left\{ {\lambda _k^{{\rm{O}}}} \right\}} \right) \\
\label{dual1 D_OMA}{\rm{s.t.}}\;\;&{\sum\nolimits_{k = 1}^K {{\alpha _k}\lambda _k^{{\rm{O}}}}  = 1},{\lambda _k^{{\rm{O}}} \ge 0},\forall k.
\end{align}
\end{subequations}
\vspace{-1.2cm}

\noindent As the strong duality holds, we can solve problem (P2) by solving its dual problem (D2). First, we solve problem \eqref{Lagrange dual function D_OMA} to obtain ${f_2}\left( {\left\{ {\lambda _k^{{\rm{O}}}} \right\}} \right)$ under any given dual variables ${\left\{ {{\lambda _k^{{\rm{O}}}}} \right\}}$. With the given dual variables, problem \eqref{Lagrange dual function D_OMA} can be decomposed into the following subproblems
\vspace{-0.4cm}
\begin{align}\label{sub1 D OMA}
\mathop {\max }\limits_{R_\infty^{{\rm{O}}}} \;\;\left( {1 - \sum\nolimits_{k = 1}^K {{\alpha _k}{\lambda _k^{{\rm{O}}}}} } \right)R_\infty^{{\rm{O}}}
\end{align}
\vspace{-1.2cm}
\begin{subequations}\label{D OMA}
\begin{align}
\mathop {\max }\limits_{\left\{ {{p_k}\left[ n \right],{\omega _k}\left[ n \right]} \right\},{\mathbf{\Theta}} \left[ n \right]} \;&\sum\nolimits_{k = 1}^K {{\varphi _k}\left( {{\mathbf{\Theta}} \left[ n \right],{p_k}\left[ n \right],{\omega _k}\left[ n \right],\lambda _k^{{\rm{O}}}} \right)} ,\forall n  \\
\label{con D OMA}{\rm{s.t.}}\;\;&\eqref{discrete phase shift OMA}-\eqref{resource allocation2 OMA},
\end{align}
\end{subequations}
\vspace{-1.2cm}

\noindent where ${\varphi _k}\left( {{\mathbf{\Theta}} \left[ n \right],{p_k}\left[ n \right],{\omega _k}\left[ n \right],\lambda _k^{{\rm{O}}}} \right) = \frac{{\lambda _k^{{\rm{O}}}}}{T}{\omega _k}\left[ n \right]{\log _2}\left( {1 + \frac{{{{\left| {{h_k} + {\mathbf{g}}_k^H{\mathbf{\Theta}} \left[ n \right]{\mathbf{v}}} \right|}^2}{p_k}\left[ n \right]}}{{{\omega _k}\left[ n \right]{\sigma ^2}}}} \right)$.\\
\indent As $\sum\nolimits_{k = 1}^K {{\alpha _k}{\lambda _k^{{\rm{O}}}}}  = 1$, the objective function value of subproblem \eqref{sub1 D OMA} is always zero. In this case, we can choose any arbitrary real number as the optimal solution $R_\infty^{*{{\rm{O}}}}$. We set $R_\infty^{*{{\rm{O}}}}=0$ for simplicity. Therefore, we just need to focus on subproblem \eqref{D OMA}. Since the subproblems in \eqref{D OMA} are identical for different time blocks $n$'s, we can drop the index $n$ for ease of exposition. We denote the optimal solutions to problem \eqref{D OMA} as ${{\mathbf{\Theta}} ^*}$, $\left\{ {p_k^*} \right\}$ and $\left\{ \omega _k^* \right\}$. To solve problem \eqref{D OMA}, we have the following lemma.
\vspace{-0.4cm}
\begin{lemma}\label{D OMA 1}
\emph{The optimal IRS reflection matrix, power allocation and orthogonal resource allocation to problem \eqref{D OMA} are given by}
\vspace{-0.4cm}
\[{{\mathbf{\Theta}} ^*} = {{\mathbf{\Theta}} _{{k^*}}},\;p_k^* = \left\{ \begin{gathered}
  {P_{\max }},{\rm{if}}\;\;k = {k^*} \hfill \\
  0,\;{\rm {otherwise}} \hfill \\
\end{gathered}  \right.,\omega _k^* = \left\{ \begin{gathered}
  1,{\rm{if}}\;\;k = {k^*} \hfill \\
  0,\;{\rm {otherwise}} \hfill \\
\end{gathered}  \right.\]
\vspace{-0.6cm}

\noindent \emph{where ${{\mathbf{\Theta}} _k} = \arg \mathop {\max }\limits_{{\mathbf{\Theta}}  \in {\mathcal{S}}} \;\;{\left| {{h_k} + {\mathbf{g}}_k^H{\mathbf{\Theta}} {\mathbf{v}}} \right|^2},\forall k$ and ${k^*} = \arg \mathop {\max }\limits_{k \in {\mathcal{K}}} \;\frac{{\lambda _k^{{\rm{O}}}}}{T}{\log _2}\left( {1 + \frac{{{{\left| {{h_k} + {\mathbf{g}}_k^H{\mathbf{\Theta}} _k {\mathbf{v}}} \right|}^2}{P_{\max }}}}{{{\sigma ^2}}}} \right)$.}
\begin{proof}
See Appendix C.
\end{proof}
\end{lemma}
\vspace{-0.3cm}
Similarly, the optimal solution $R_\infty^{*{{\rm{O}}}}$ is generally non-unique since $\sum\nolimits_{k = 1}^K {{\alpha _k}{\lambda _k^{{\rm{O}}}}}  = 1$. Additional steps are required to construct the optimal primal solution to problem (P2). Furthermore, \textbf{Lemma \ref{D OMA 1}} reveals that there is only one user served according to the optimal solution to problem \eqref{D OMA}. With this insight, the total non-unique optimal solutions ${{\mathbf{\Theta}} ^*}$, $\left\{ {p_k^*} \right\}$ and $\left\{ \omega _k^* \right\}$ to problem \eqref{D OMA} can be directly obtained using the following proposition instead of finding the optimal dual solutions ${\left\{ {{\lambda _k^{*{\rm{O}}}}} \right\}}$ as did in the previous section.
\vspace{-0.4cm}
\begin{proposition}\label{optimal Alternative transmission}
\emph{For a given rate-profile vector ${\bm \alpha}$, let $\Upsilon $ denote the user index set with a non-zero rate target ratio, $\Upsilon  = \left\{ {k|{\alpha _k} > 0} \right\}$. Suppose that the optimal dual solutions are ${\left\{ {\lambda _k^{*{\rm{O}}}} \right\}}$ to (D2), then problem \eqref{D OMA} has a total of $\left| \Upsilon  \right|$ optimal solutions $\left\{ {{\Gamma _k},{k \in \Upsilon }} \right\}$ which are given by}
\vspace{-0.3cm}
\begin{align}\label{K optimal D OMA}
{\Gamma _k} = \left\{ {{{\mathbf{\Theta}} _k},\left( {{{\mathbf{0}}_{k - 1}},{P_{\max }},{{\mathbf{0}}_{K - k}}} \right),\left( {{{\mathbf{0}}_{k - 1}},1,{{\mathbf{0}}_{K - k}}} \right)} \right\},
\end{align}
\vspace{-1cm}

\noindent \emph{where ${{\mathbf{\Theta}} _k} = \arg \mathop {\max }\limits_{{\mathbf{\Theta}}  \in {\mathcal{S}}} \;\;{\left| {{h_k} + {\mathbf{g}}_k^H{\mathbf{\Theta}} {\mathbf{v}}} \right|^2},{k \in \Upsilon }$ and it must hold that ${\varphi _k}\left( {{\Gamma _k},\lambda _k^{*{\rm{O}}}} \right) = {\varphi _i}\left( {{\Gamma _i},\lambda _i^{*{\rm{O}}}} \right),$ \\$ {k,i \in \Upsilon }$ for problem \eqref{D OMA}.}
\begin{proof}
\emph{This is shown by contradiction. Suppose that the $k$th term ${\varphi _k}\left( {{\Gamma _k},\lambda _k^{*{\rm{O}}}} \right)$ is smaller than any one of the other $\left| \Upsilon  \right| - 1$ terms (i.e., ${\varphi _i}\left( {{\Gamma _i},\lambda _i^{*{\rm{O}}}} \right),\forall i \in \Upsilon ,i \ne k$). In this case, ${\text{}}{\Gamma _k}$ cannot be the optimal solution to problem \eqref{D OMA}. Then, the $k$th user cannot be served throughout the whole period $T$, which causes a zero rate for the $k$th user with a non-zero rate requirement. As a result, to achieve a non-zero rate, it must hold that ${\varphi _k}\left( {{\Gamma _k},\lambda _k^{*{\rm{O}}}} \right) = {\varphi _i}\left( {{\Gamma _i},\lambda _i^{*{\rm{O}}}} \right),{k,i \in \Upsilon }$, which contradicts our initial assumption and the proposition is proved.}
\end{proof}
\end{proposition}
\vspace{-0.4cm}
Based on \textbf{Proposition \ref{optimal Alternative transmission}}, we need to determine the time-sharing ratio among the $\left| \Upsilon  \right|$ optimal solutions $\left\{ {{\Gamma _k},{k \in \Upsilon }} \right\}$ to construct the optimal primal solution to problem (P2). Here, time-sharing means that the total $K$ users should be served in an alternating manner for a certain portion of the total block duration $T$. Let ${\tau _k}$ denote the optimal transmission duration for the $k$th user. Then, the optimal primal solution to (P2) can be obtained by solving the following problem
\vspace{-0.6cm}
\begin{subequations}\label{P1-OMA}
\begin{align}
&\mathop {\max }\limits_{R_\infty^{{\rm{O}}},\left\{ {{\tau _k} \ge 0} \right\}} \;\;R_\infty^{{\rm{O}}} \\
\label{P1-OMA1}{\rm{s.t.}}\;\;\frac{{{\tau _k}}}{T}{\log _2}&\left( {1 + \frac{{{{\left| {{h_k} + {\mathbf{g}}_k^H{{\mathbf{\Theta}} _k}{\mathbf{v}}} \right|}^2}{P_{\max }}}}{{{\sigma ^2}}}} \right) \ge {\alpha _k}R_\infty^{{\rm{O}}},\forall k,\\
\label{P1-OMA2}&\sum\nolimits_{k = 1}^{\left| \Upsilon  \right|} {{\tau _k}}  = T.
\end{align}
\end{subequations}
\vspace{-0.6cm}

\noindent It can be verified that the above problem is a standard LP, which can be solved by using standard convex optimization tools such as CVX \cite{cvx}. Therefore, the optimal solution to (P2) with a given rate-profile vector ${\bm \alpha}$ can be obtained. The algorithm for optimally solving problem (P2) is summarized in \textbf{Algorithm 2}. The complexity of step 1) is ${\mathcal{O}}\left( {{L^M}{\left| \Upsilon  \right|}} \right)$ and of solving the LP problem \eqref{P1-OMA} is ${\mathcal{O}}\left( { {{\left| \Pi  \right|}^3}} \right)$ \cite{convex}. The total complexity of Algorithm 2 is ${\mathcal{O}}\left( {{L^M}{\left| \Upsilon  \right|} + {{\left| \Upsilon  \right|}^3}} \right)$.
\begin{algorithm}[!t]\label{method2}
\caption{Algorithm for Optimally Solving Problem (P2) when $N \to \infty $}
\begin{algorithmic}[1]
\STATE Find the total $\left| \Upsilon  \right|$ optimal solutions $\left\{ {{\Gamma _k},{k \in \Upsilon }} \right\}$ with \eqref{K optimal D OMA}.
\STATE Obtain the optimal solution $R_\infty ^{*{{\rm{O}}}}$ to problem (P2) via time-sharing by solving problem \eqref{P1-OMA}.
\end{algorithmic}
\end{algorithm}
\vspace{-0.3cm}
\begin{remark}\label{Alternative transmission}
\emph{The optimal solution to problem \eqref{P1-OMA} unveils that to achieve any point on the Pareto boundary of the rate region ${{\mathcal{R}}}\left( {b,\infty } \right)$ in the OMA scheme, the optimal transmission strategy is alternating transmission among each individual user with its combined channel power gain maximized by dynamically reconfiguring the IRS reflection matrix.}
\end{remark}
\vspace{-0.6cm}
\begin{remark}\label{upper bound}
\emph{If the IRS is equipped with continuous phase shifts, the closed-form solution to ${{\mathbf{\Theta}} _k} = \arg \mathop {\max }\limits_{{\mathbf{\Theta}}  \in {\mathcal{S}}} \;\;{\left| {{h_k} + {\mathbf{g}}_k^H{\mathbf{\Theta}} {\mathbf{v}}} \right|^2}$ is $\theta _m^{*k} = \arg \left( {{h_k}} \right) - \arg \left( {g_{m,k}^H{v_m}} \right)$, where ${g_{m,k}^H}$ and ${{v_m}}$ are the $m$th element of ${{\mathbf{g}}_k^H}$ and ${\mathbf{v}}$, respectively. This closed form solution follows intuitively from: 1) Triangle Inequality which says that the magnitude of the sum of 2 complex vectors is maximized when the 2 vectors are aligned (same direction). In this case: $\left| {\mathbf{x} + \mathbf{y}} \right| = \left| \mathbf{x} \right| + \left| \mathbf{y} \right|$. 2) Cauch-Schwartz Inequality which says that the magnitude of the dot product is maximized when the two vectors are aligned. The rate region achieved with continuous phase shifts in OMA provides an upper bound to that with discrete phase shifts.}
\end{remark}
\vspace{-1cm}
\subsection{Rate Region Inner Bound with Finite $N$}
\vspace{-0.3cm}
In this subsection, we derive an inner bound of the rate region ${{{\mathcal{R}}}}\left( {b,N} \right)$ with finite value $N$. Similarly, based on the obtained optimal solutions $\left\{ {{\mathbf{\Theta}} _k^*,{\tau _k}} \right\}_{k = 1}^\Upsilon $ in the previous subsection, the IRS reflection matrix $\left\{ {{\mathbf{\Theta}} \left[ n \right]} \right\}_{n = 1}^{{{N}}}$ in the OMA transmission scheme over finite $N$ time blocks is given by
\vspace{-0.4cm}
\begin{align}
\left\{ {{\mathbf{\Theta}} \left[ n \right] = {\mathbf{\Theta}} _k^*,{N_{k - 1}} + 1 \le n \le {N_k}} \right\}_{k = 1}^\Upsilon,
\end{align}
\vspace{-1.2cm}

\noindent where ${N_k} = \left[ {\frac{{{T_k}}}{T}N} \right]$, ${T_k} = \sum\nolimits_{i = 0}^k {{\tau _i}} $ and ${{\tau _0} = 0}$.\\
\indent Next, under the designed IRS reflection matrix $\left\{ {{\mathbf{\Theta}} \left[ n \right]} \right\}_{n = 1}^{{{N}}}$, problem (P2) can be written as the following resource allocation problem
\vspace{-0.5cm}
\begin{subequations}\label{P1 OMA sub1}
\begin{align}
&\mathop {\max }\limits_{R^{\rm{O}},\left\{ {{p_k}\left[ n \right],{\omega _k}\left[ n \right]} \right\}} \;\;R^{\rm{O}}  \\
\label{rate allocation OMA sub}{\rm{s.t.}}\;\;&\frac{1}{N}\sum\nolimits_{n = 1}^{{{N}}} {{\omega _k}\left[ n \right]{{\log }_2}\left( {1 + \frac{{{{\left| {{h_k} + {\mathbf{g}}_k^H{\mathbf{\Theta}} \left[ n \right]{\mathbf{v}}} \right|}^2}{p_k}\left[ n \right]}}{{{\omega _k}\left[ n \right]{\sigma ^2}}}} \right)}  \ge {\alpha _k}{R^{\rm{O}}},\\
\label{total power1 OMA sub}&\sum\nolimits_{k = 1}^K { {{p_k}\left[ n \right]} } \le {P_{\max}},\forall n,\\
\label{resource allocation1 OMA sub}&\sum\nolimits_{k = 1}^K {{\omega _k}} \left[ n \right] \le 1,\forall n,\\
\label{resource allocation2 OMA sub}&{p_k}\left[ n \right] \ge 0, 0 \le {\omega _k}\left[ n \right] \le 1,\forall k,n,
\end{align}
\end{subequations}
\vspace{-1.2cm}

\noindent As the left-hand-side of constraint \eqref{rate allocation OMA sub} is jointly concave with respect to ${{\omega _k}\left[ n \right]}$ and ${{p_k}\left[ n \right]}$, problem \eqref{P1 OMA sub1} is a convex problem. We can solve it by utilizing standard convex optimization techniques such as the interior point method \cite{convex}. The complexity for solving problem \eqref{P1 OMA sub1} is ${\mathcal{O}}\left( {{{\left( {2KN}+1 \right)}^{3.5}}} \right)$~\cite{convex}, where ${2KN}$ stands for the number of optimization variables of \eqref{P1 OMA sub1}. As a result, an inner bound of rate region ${{{\mathcal{R}}}}\left( {b,N} \right)$ with finite $N$ can be efficiently obtained. Similarly, the total complexity of the proposed suboptimal approach for OMA is ${\mathcal{O}}\left( {{L^M}{\left| \Upsilon  \right|} + {{\left| \Upsilon  \right|}^3 + {{{\left( {2KN} \right)}^{3.5}}}}} \right)$ due to the adoption of \textbf{Algorithm 2}. It is worth noting that the optimal solution to (P2) with finite $N$ can also be obtained by exhaustively searching all IRS reflection matrix configurations and solving the resulting resource allocation problem \eqref{P1 OMA sub1}, which serves as a baseline scheme in the next section.
\vspace{-0.6cm}
\section{Numerical Results}
\begin{figure}[t!]
    \begin{center}
        \includegraphics[width=2.8in]{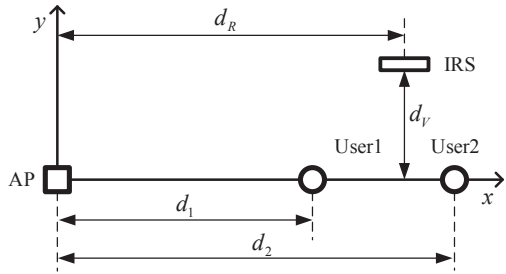}
        \caption{The simulated IRS-assisted 2-user communication scenario.}
        \label{setip}
    \end{center}
\end{figure}

\vspace{-0.3cm}
In this section, numerical examples are provided to validate our proposed designs. As illustrated in Fig. \ref{setip}, an IRS-assisted multi-user communication system is considered, in which the AP and the IRS are located at $\left( {0,0,0} \right)$ meters and $\left( {{d_R},{d_V},0} \right)$ meters, respectively. We consider the case with $K=2$ users, whose locations are set as $\left( {{d_1},0,0} \right)$ meters and $\left( {{d_2},0,0} \right)$ meters. The distances for the direct link, the AP-IRS link and the IRS-user link are denoted by ${d_{AU,k}}$, ${d_{AI}}$ and ${d_{IU,k}}$, respectively. The distance-dependent path loss for all channels is modeled as $PL\left( d \right) = {\rho _0}{\left( {\frac{d}{{{d_0}}}} \right)^{ - \alpha }}$, where ${\rho _0} =  - 30$ dB denotes the path loss at the reference distance ${d_0} = 1$ meter (${\rm{m}}$), $d$ denotes the link distance and $\alpha$ denotes the path loss exponent. We set ${d_1}=43$ m, ${d_2}=50$ m, ${d_R}=49$ m and ${d_V}=1$ m. The size of each sub-surface is set to $B=4$. For small scale fading, the Rayleigh fading channel model and the Rician fading model are assumed for the direct link and the AP-IRS/IRS-user links, respectively. Then, the corresponding channel coefficients are given by
\vspace{-0.5cm}
\begin{subequations}
\begin{align}\label{channel coefficients1}
&{{{h}}_k} = \sqrt {PL\left( {{d_{AU,k}}} \right)} {{h}}_k^{{\rm{NLoS}}},k \in {\mathcal{K}},\\
\label{channel coefficients2}&{\mathbf{v}} = \sqrt {\frac{{PL\left( {{d_{AI}}} \right)}}{{{K_{AI}} + 1}}} \left( {\sqrt {{K_{AI}}} {\mathbf{v}}^{{\rm{LoS}}} + {\mathbf{v}}^{{\rm{NLoS}}}} \right),\\
\label{channel coefficients3}&{{\mathbf{g}}_k} = \sqrt {\frac{{PL\left( {{d_{IU,k}}} \right)}}{{{K_{IU}} + 1}}} \left( {\sqrt {{K_{IU}}} {\mathbf{g}}_k^{{\rm{LoS}}} + {\mathbf{g}}_k^{{\rm{NLoS}}}} \right),k \in {\mathcal{K}},
\end{align}
\end{subequations}
\vspace{-1.0cm}

\noindent where ${{K_{AI}}}$ and ${{K_{IU}}}$ denote the Rician factors of the AP-IRS/IRS-user links. ${{\mathbf{v}}^{\rm{LoS}}}$ and ${{\mathbf{g}}_k^{\rm{LoS}}}$ denote the deterministic LoS components, ${{h}}_k^{{\rm{NLoS}}}$, ${{\mathbf{v}}^{{\rm{NLoS}}}}$ and ${{\mathbf{g}}_k^{{\rm{NLoS}}}}$ denote the Rayleigh fading components. In this paper, the path loss exponents for the direct link, AP-IRS link and IRS-user link are set to be ${\alpha _{AU}} = 3.5$, ${\alpha _{AI}} = 2.2$ and ${\alpha _{IU}} = 2.8$, respectively\footnote{Under the considered simulation setup, the pathloss of the AP-user2 link is -89.46 dB, while the pathloss of the AP-IRS-user2 link is -102.4 dB. It can be observed that the reflection link suffers much more severe pathloss due to the ``double fading'' effect.}, the Rician factors are ${K_{AI}} = {K_{IU}} = 3$ dB, and the noise power is set to be $\sigma  =  - 80$ dBm~\cite{Wu2019IRS,Yang_OFDMA}.
\vspace{-0.6cm}
\subsection{Baseline Scheme}
\vspace{-0.2cm}
Note that for the case with finite $N$, the proposed approaches in Section III-B and Section VI-B provide an inner bound on the capacity and rate regions for NOMA and OMA. To verify the optimality of the proposed suboptimal approaches, we consider the following baseline scheme: All possible IRS reflection matrix configurations over $N$ time slots are exhaustively considered, and the resulting resource allocation problems can be solved as problems \eqref{P1 sub ap} and \eqref{P1 OMA sub1}. The complexities and achieved performances of all proposed schemes and the baseline scheme are compared in Table I. In particular, since the iterative algorithm for \eqref{P1 sub ap} only obtains a suboptimal solution, the baseline scheme for NOMA is also suboptimal. However, for OMA, the baseline scheme is capable of obtaining a globally optimal solution since the resulting resource allocation problem \eqref{P1 OMA sub1} is a convex optimization problem which has a globally optimal solution.
\begin{table}[!t]\footnotesize
\caption{Computational Complexity and Performance of Proposed and Baseline Schemes.}
\begin{center}
\centering
\resizebox{\textwidth}{!}{
\begin{tabular}{|l|l|l|l|l|l|}
\hline
\centering
 \makecell[c]{\textbf{{MA}}} &\makecell[c]{\textbf{{Scheme}}}  &\textbf{\makecell[c]{IRS reflection\\ matrix configuration}} &\textbf{\makecell[c]{Resource\\ Allocation}} & \makecell[c]{\textbf{{Computational Complexity}}} &\makecell[c]{\textbf{{Performance}}}\\
\hline
\centering
\multirow{3}{*}{\makecell[c]{NOMA}} &\makecell[c]{$N \to \infty $,\\ proposed scheme} &\multicolumn{2}{c|}{\makecell[c]{Algorithm 1}} &{${\mathcal{O}}\left( {\frac{{{L^M}{K^4}}}{\varepsilon } + {{\left| \Pi  \right|}^3}} \right)$} & {Optimal}\\
\cline{2-6}
\centering
 &\makecell[c]{Finite $N$,\\ proposed scheme}&\makecell[c]{Reconstruct from\\ Algorithm 1}&\makecell[c]{Iterative\\ solve \eqref{P1 sub ap}} & {${{\mathcal{O}}}\left( {\frac{{{L^M}{K^4}}}{\varepsilon } + {{\left| \Pi  \right|}^3} + I{{\left( {2{K^2}N + 1} \right)}^{3.5}}} \right)$} & {Suboptimal}\\
\cline{2-6}
\centering
 &\makecell[c]{Finite $N$,\\ baseline scheme} &\makecell[c]{Exhaustive\\ Search}&\makecell[c]{Iterative\\ solve \eqref{P1 sub ap}} &${\mathcal{O}}\left( {{L^{MN}}\left( {I{{\left( {2{K^2}N + 1} \right)}^{3.5}}} \right)} \right)$ & {Suboptimal}\\
\hline
\centering
\multirow{3}{*}{\makecell[c]{OMA}}  &\makecell[c]{$N \to \infty $,\\ proposed scheme} &\multicolumn{2}{c|}{\makecell[c]{Algorithm 2}} &{${\mathcal{O}}\left( {{L^M}{\left| \Upsilon  \right|} + {{\left| \Upsilon  \right|}^3}} \right)$ }&Optimal\\
\cline{2-6}
\centering
 &\makecell[c]{Finite $N$,\\ proposed scheme}&\makecell[c]{Reconstruct from\\ Algorithm 2}&\makecell[c]{Solve \eqref{P1 OMA sub1}}  & ${\mathcal{O}}\left( {{L^M}{\left| \Upsilon  \right|} + {{\left| \Upsilon  \right|}^3 + {{{\left( {2KN} \right)}^{3.5}}}}} \right)$ &Suboptimal\\
\cline{2-6}
\centering
 &\makecell[c]{Finite $N$,\\ baseline scheme} &\makecell[c]{Exhaustive\\ Search}&\makecell[c]{Solve \eqref{P1 OMA sub1}} &${\mathcal{O}}\left( {{L^{MN}}{{\left( {2KN} \right)}^{3.5}}} \right)$ &Optimal\\
\hline
\end{tabular}
}
\end{center}
\label{comparison}
\end{table}
\vspace{-0.6cm}
\subsection{Capacity and Rate Regions of IRS for $N \to \infty $}
\begin{figure}[t!]
\centering
\subfigure[Capacity regions with NOMA.]{\label{CR1}
\includegraphics[width= 2.8in]{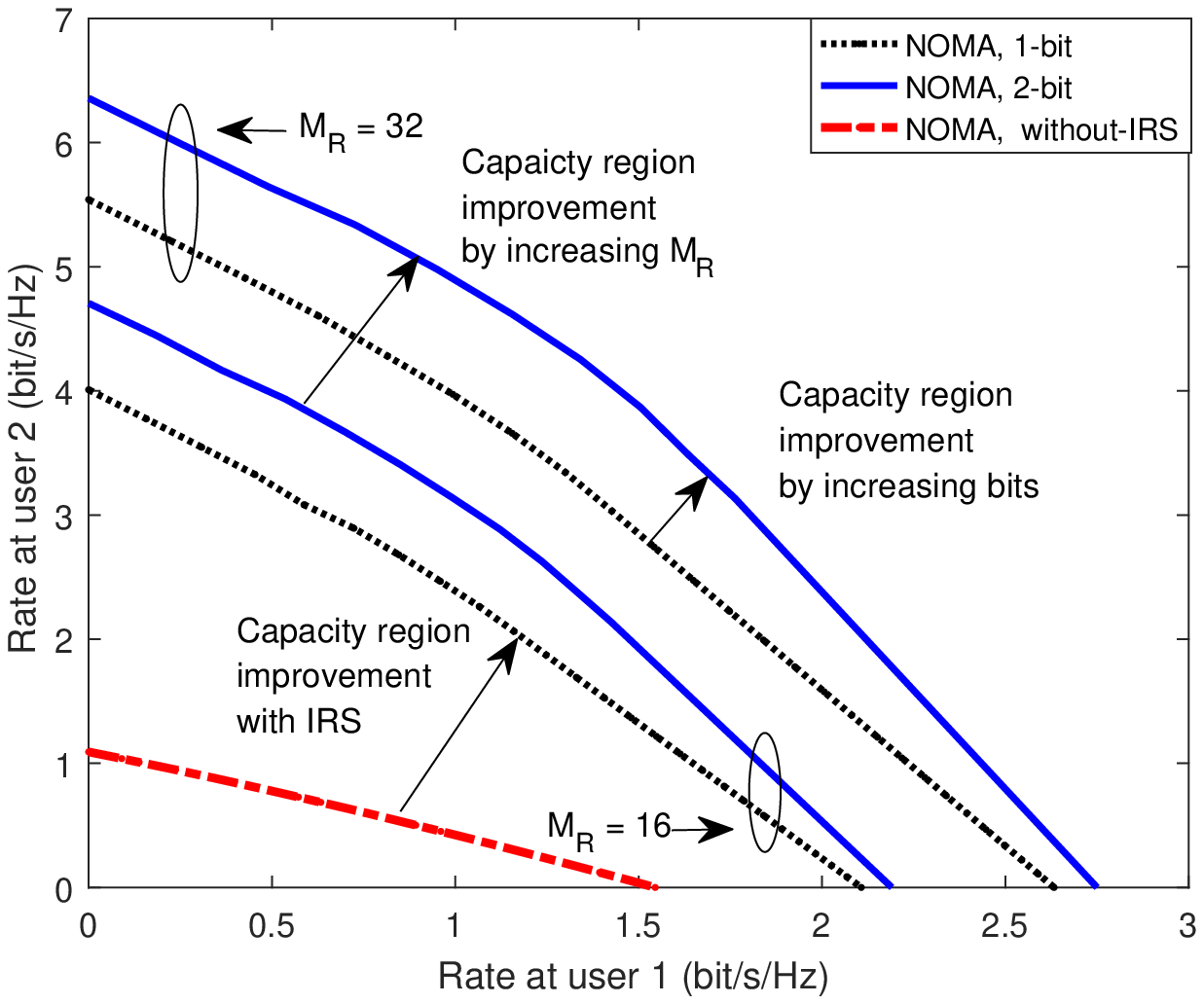}}
\subfigure[Rate regions with OMA.]{\label{RR1}
\includegraphics[width= 2.8in]{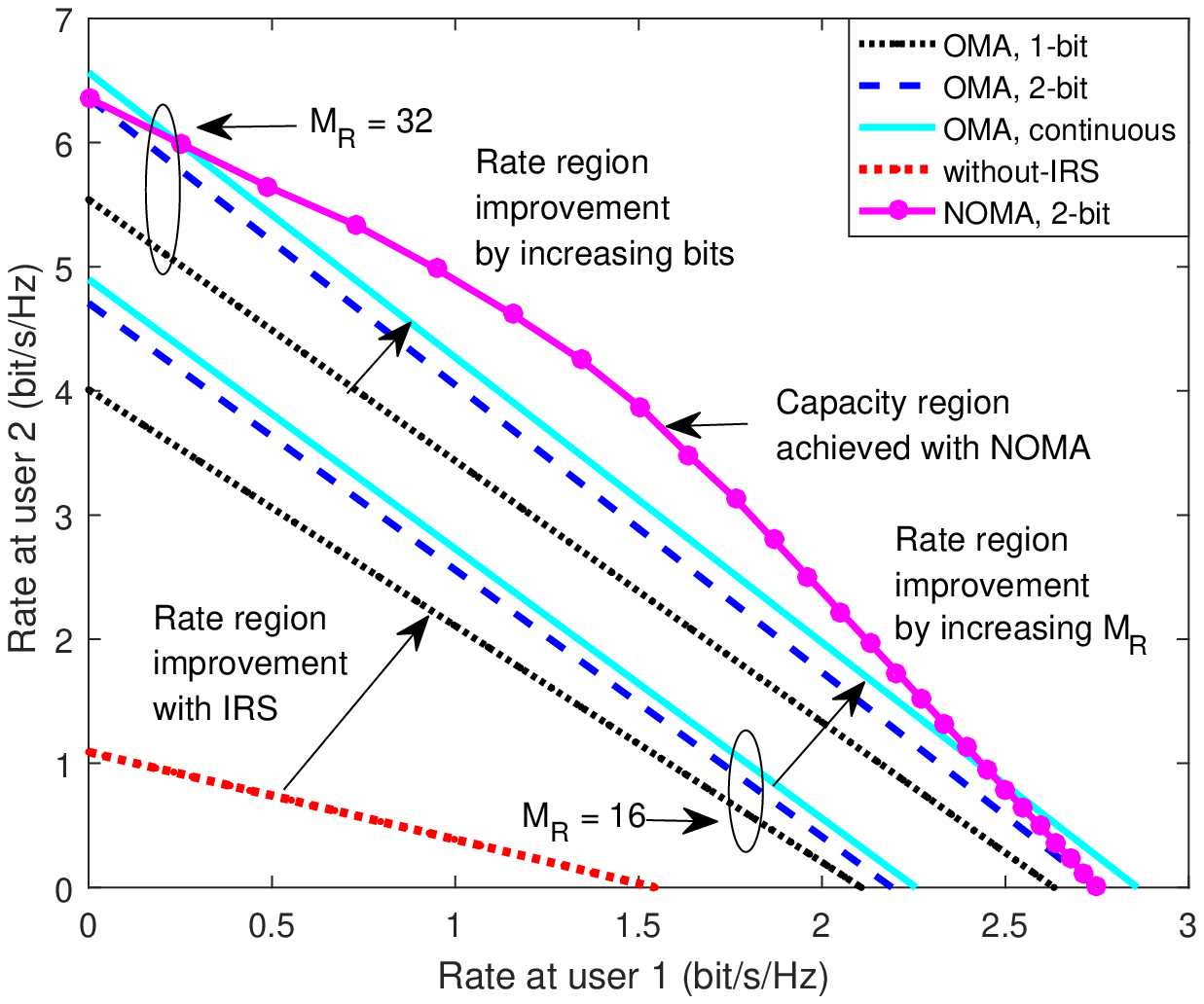}}
\setlength{\abovecaptionskip}{-0cm}
\caption{Capacity and rate regions for $N \to \infty $ and a random realization of ${h_k},{\mathbf{v}},{{\mathbf{g}}_k}$ with $P_{\max}=10$ dBm.}\label{CRR}
\end{figure}

\vspace{-0.2cm}
\indent In Fig. \ref{CR1} and Fig. \ref{RR1}, we present the capacity and rate regions achieved by \textbf{Algorithm 1} and \textbf{Algorithm 2} in the ideal case of $N \to \infty $ for different numbers of IRS reflecting elements $M_R$ and phase resolution bits $b$. The transmit power is set to $P_{\max}=10$ dBm. As illustrated in Fig. \ref{CR1}, we also provide the capacity region achieved without the IRS. It is first observed that the capacity region with the IRS is significantly larger than that without the IRS, which demonstrates the IRS performance advantages. Moreover, it is also observed that the capacity region can be improved by increasing the number of IRS reflecting elements $M_R$ because a higher array gain is achieved. For the same $M_R$, the capacity region is further enlarged by increasing the phase resolution bits $b$. This is expected since a larger $b$ leads to a more accurate IRS reflection matrix. In Fig. \ref{RR1}, the rate region of OMA with the continuous IRS phase shifts using the method in \textbf{Remark \ref{upper bound}}, the rate region achieved without the IRS and the capacity region of NOMA with 2-bit phase shifts and 32 reflecting elements are provided for comparison. Similarly, considerable rate region improvement can be achieved by the IRS with a larger number of IRS reflecting elements $M_R$ and phase resolution bits $b$. The capacity region achieved by NOMA contains the rate region of OMA for the same number of phase shifts and reflecting elements. This is expected since NOMA is a capacity-achieving transmission scheme from the perspective of information theory, while OMA is suboptimal. It can be also observed that the performance gap between the continuous phase shifts and the 2-bit phase shifts is small, which implies that the 2-bit phase shifts may serve as a promising candidate to achieve a desirable performance-complexity tradeoff. In addition, in both Fig. \ref{CR1} and Fig. \ref{RR1}, the performance enhancement of user 2 is more pronounced than that of user 1. This is because the IRS is deployed closer to user 2 in the simulation setup and its reflection link suffers less path loss than user 1.
\vspace{-0.6cm}
\subsection{Capacity and Rate Region Inner Bounds of IRS for finite $N$}
\vspace{-0.2cm}
\begin{figure}[t!]
\centering
\subfigure[Capacity region inner bounds with NOMA.]{\label{CRN}
\includegraphics[width= 2.8in]{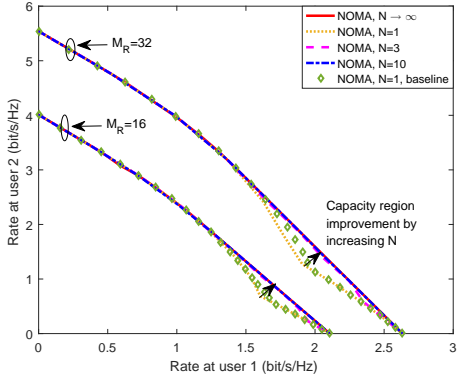}}
\subfigure[Rate region inner bounds with OMA.]{\label{RRN}
\includegraphics[width= 2.8in]{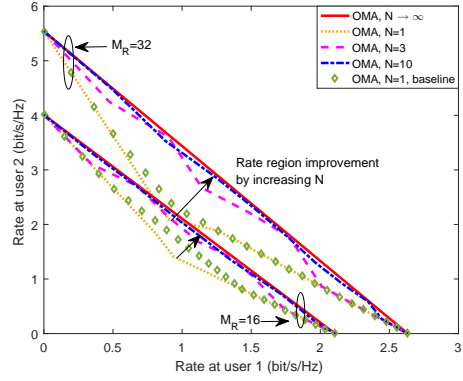}}
\setlength{\abovecaptionskip}{-0cm}
\caption{Capacity and rate region inner bounds for finite $N$ and a random realization of ${h_k},{\mathbf{v}},{{\mathbf{g}}_k}$ with $P_{\max}=10$ dBm and $b=1$.}\label{CRRN}
\end{figure}

In Fig. \ref{CRN} and Fig. \ref{RRN}, we present the proposed capacity and rate region inner bounds for NOMA and OMA in the case of finite $N$. We set $P_{\max}=10$ dBm and $b=1$. For comparison, the corresponding capacity and rate regions achieved by \textbf{Algorithm 1} and \textbf{Algorithm 2} when $N \to \infty $ and achieved by the baseline scheme when $N=1$ are also provided. As illustrated in both figures, the proposed inner bounds approach the corresponding capacity and rate regions when $N$ increases, which underscores the importance of dynamically reconfiguring the IRS reflection matrix. It is also observed that the performance loss caused by finite $N$ is more pronounced for OMA than NOMA. The performance loss becomes negligible for NOMA with only $N=3$, while it requires $N=10$ for OMA. This interesting insight unveils the advantages of NOMA transmission in IRS-assisted networks, since NOMA not only achieves a higher capacity but also requires less hardware complexity for real-time IRS control. For the baseline scheme, we only provide the results for the case of $N=1$ since the computational complexity increases exponentially\footnote{As presented in Table I, considering the case of $N=3$, $M=8$ and $b=1$, the baseline scheme needs to search ${{2^{8 \times 3}}}=16777216$ combinations of the IRS reflection configuration. However, the computational complexity of the proposed scheme is on the order of ${2^8} = 256$, which is significantly lower than the baseline scheme.} with the increase of $N$. It can be observed that there is a slight performance gap between the proposed scheme and the baseline scheme with $N=1$. However, the proposed scheme has a much lower computational complexity than the baseline scheme and achieves a near-optimal performance.
\vspace{-0.6cm}
\subsection{Common Average Rate Performance}
\begin{figure}[t!]
\centering
\begin{minipage}[b]{0.48\linewidth}
\includegraphics[width=2.8in]{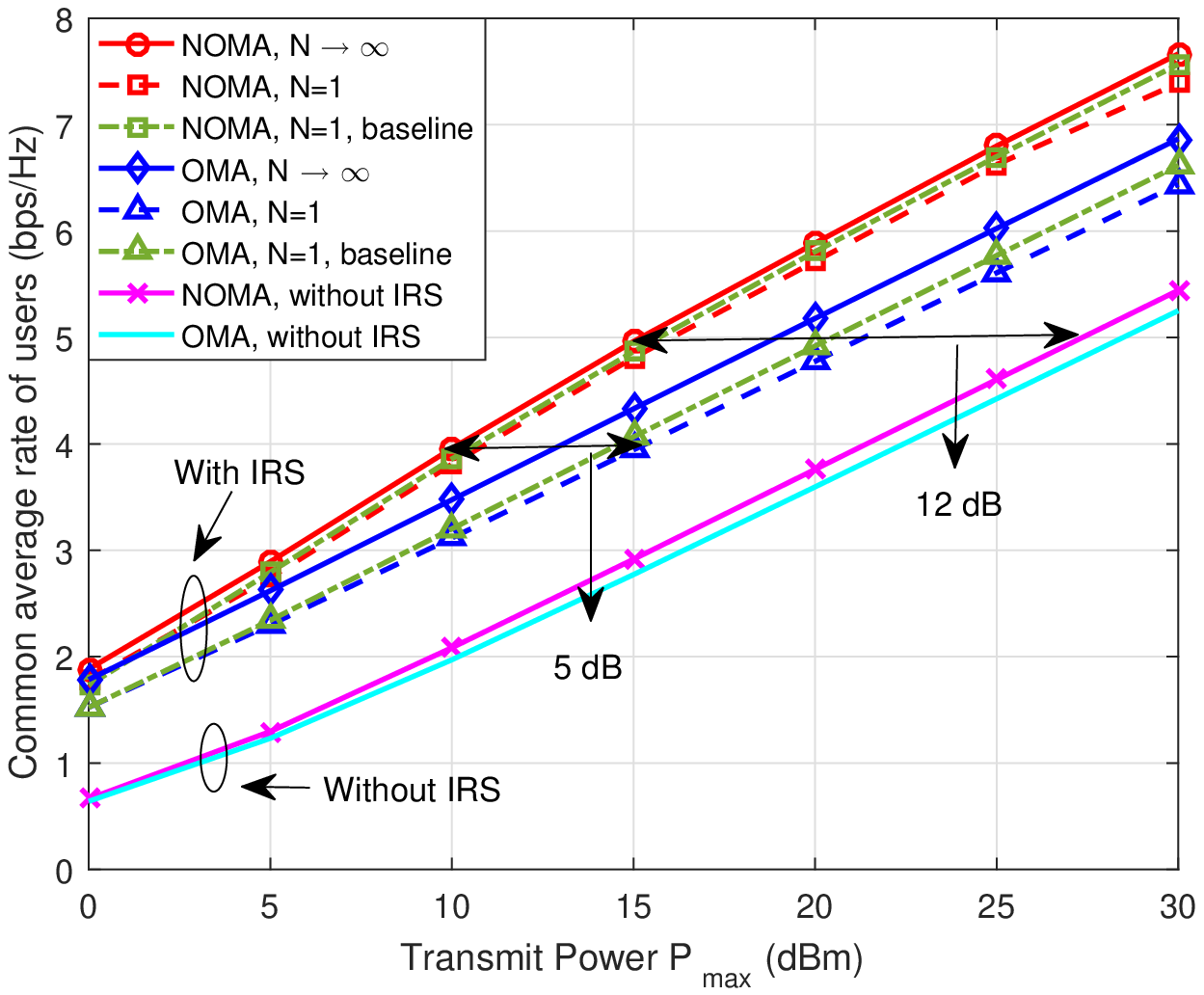}
\caption{The common average rate versus $P_{\max}$ for $M_R=32$ and $b=1$.}
\label{SRvP}
\end{minipage}
\quad
\begin{minipage}[b]{0.48\linewidth}
\includegraphics[width=2.8in]{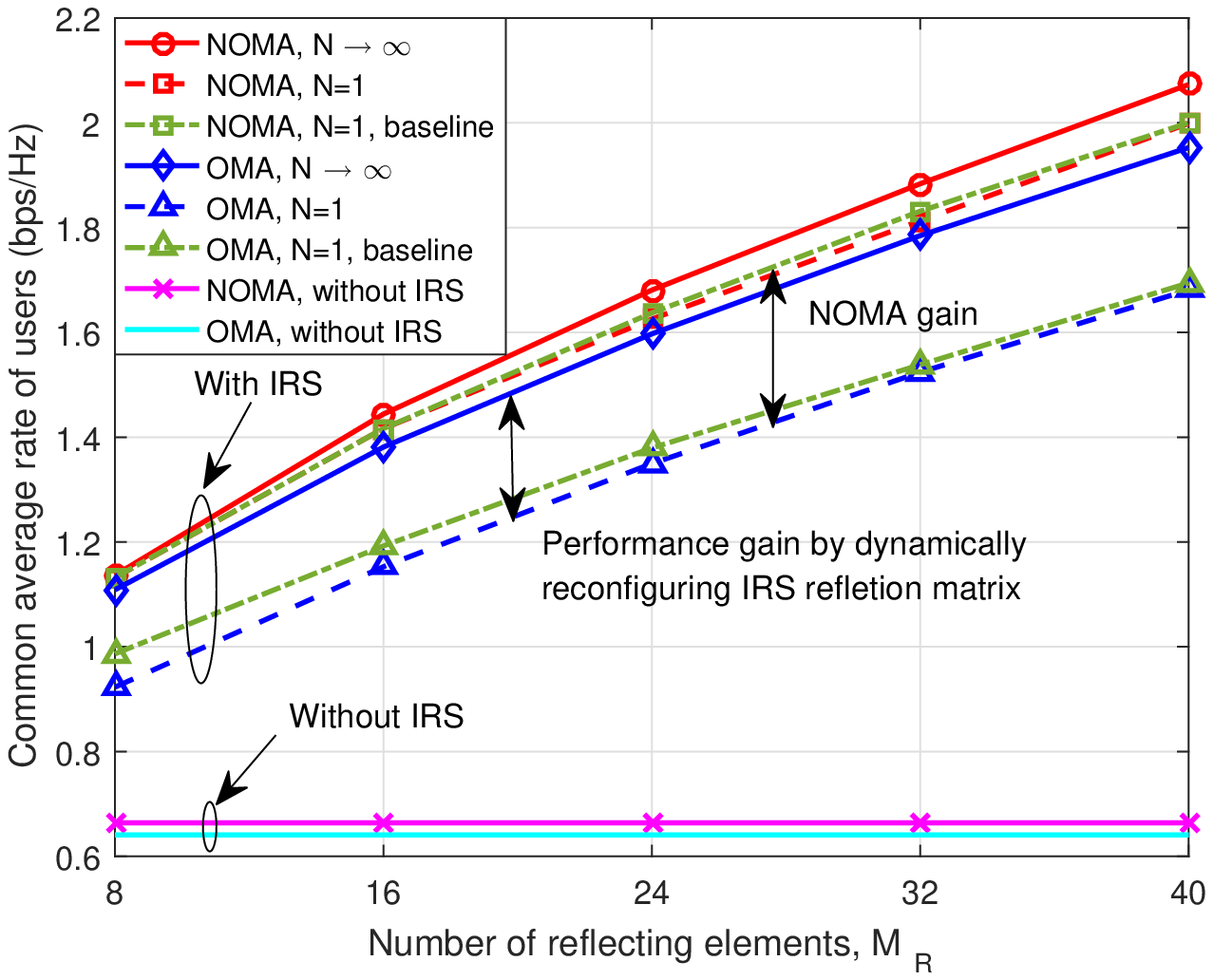}
\caption{The common average rate versus $M_R$ for $P_{\max}=10$ dBm and $b=1$.}
\label{SRvM}
\end{minipage}
\end{figure}

\vspace{-0.2cm}
In this subsection, we set ${\alpha _1} = {\alpha _2} = 0.5$ and present the common average data rate performances. We consider the following schemes
\begin{itemize}
  \item \textbf{$N \to \infty $}: This is the ideal case, where the IRS reflection matrix can be configured in a real-time manner. The common average data rate is obtained using Algorithms 1 and 2 for NOMA and OMA, respectively.
  \item \textbf{$N = 1$}: In this case, the IRS reflection matrix is fixed throughout the entire data transmission. The common average data rate is obtained with our proposed inner bound designs by setting $N=1$ for NOMA and OMA.
  \item \textbf{$N = 1$, baseline}: In this case, the common average data rate is obtained with the baseline scheme by exhaustively searching over all possible IRS reflection matrix configurations for $N=1$ and solving the remaining resource allocation problem for NOMA and OMA, as presented in Table I.
  \item \textbf{without IRS}: In this case, the AP serves two users without the aid of IRS. The common average data rate is obtained by solving a conventional resource allocation problem for NOMA and OMA.
\end{itemize}
All results in Fig. \ref{SRvP} and Fig. \ref{SRvM} are averaged over 100 independent channel realizations.
\subsubsection{Common Average Rate versus Transmit Power $P_{\max}$} Fig. \ref{SRvP} shows the common average rate versus the maximum transmit power $P_{\max}$ for different schemes and $M_R=32$, $b=1$. It is observed that for all schemes, the sum rate performances increase with $P_{\max}$. Our proposed IRS schemes significantly outperform the scheme without the IRS. To achieve an identical common average rate, the IRS-assisted schemes require much less transmit power. A 12 dB performance gain can be achieved by the IRS-assisted NOMA scheme over the scheme without the IRS. Furthermore, NOMA is capable of achieving a higher performance than OMA in both $N \to \infty $ and $N =1$ cases. A 5 dB performance gain can be achieved by NOMA over OMA for $N =1$. In particular, the schemes for $N \to \infty $ achieve the best performance. This is expected since dynamically configuring the IRS reflection matrix increases the DoF to enhance the performance. This also validates the importance of designing real-time IRS control link. It is also observed that the performance loss between the baseline scheme and the proposed scheme is negligible for both NOMA and OMA with $N=1$, which is consistent with the results in Fig. \ref{CRRN}. This also verifies the effectiveness and optimality of the proposed suboptimal approaches.
\subsubsection{Common Average Rate versus the Number of IRS Elements $M_R$} Fig. \ref{SRvM} depicts the common average rate versus the number of IRS reflecting elements $M_R$ for different schemes and $P_{\max}=10$ dBm, $b=1$. For all IRS-assisted schemes, as $M_R$ increases, the achieved common average rate increases, while the performance of the scheme without the IRS remains unchanged. The performance gain achieved by reconfiguring the IRS reflection matrix is more pronounced for OMA. Similarly, the baseline scheme only slightly outperforms the proposed suboptimal approaches for NOMA and OMA. It is worth pointing out that the performance gain of NOMA over OMA in the proposed IRS-assisted scheme is more noticeable than that in the scheme without the IRS. This is because the IRS is capable of enlarging the channel power gain disparity among users, where NOMA can achieve a higher performance gain than OMA.
\vspace{-0.6cm}
\section{Conclusions and Future Work}
\vspace{-0.2cm}
In this paper, the fundamental capacity limits of IRS-assisted multi-user wireless communications were investigated. The IRS reflection matrix and resource allocation were jointly optimized for characterizing the Pareto boundary of the capacity and rate regions for NOMA and OMA transmission schemes, under the constraints of discrete phase shifts and a finite number of IRS reconfiguration times. For each scheme, the globally optimal solution was firstly obtained using the Lagrange duality method for the ideal case with an asymptotically large number of IRS reconfiguration times. It is shown that the optimal transmission strategy for NOMA is alternating transmission among different user groups and decoding orders, while for OMA it is within each individual user. Based on these solutions, the inner bounds of capacity and rate regions are efficiently derived for the general case with a finite number of IRS reconfiguration times. Numerical results showed that significant capacity gains can be achieved by deploying the IRS and revealed the importance of designing a real-time IRS control link. Due to the space limitation, some other important issues remain unaddressed in this work, which are discussed below to motivate future work.
\begin{itemize}
  \item \emph{Ergodic capacity characterization over fading channels:} This paper considered a static or quasi-static scenario and focused on one specific channel coherence time block $T$, which can be relatively large for reconfiguring the IRS multiple times, i.e., $T \gg \delta $. Such dynamic IRS reconfiguration may not hold in a high-speed mobile scenario, where the value of $T$ is comparable with or even less than $\delta $. In this case, the involved channels can be modeled by several fading states as $\left\{ {{h_k}\left[ i \right],{\mathbf{g}}_k^H\left[ i \right],{\mathbf{v}}\left[ i \right],i \in {{\mathcal{I}}}} \right\}$, where $i$ represents each fading state. The effective channel of a user over different fading states can be expressed as ${h_k}\left[ i \right] + {\mathbf{g}}_k^H\left[ i \right]{\mathbf{\Theta }}\left[ i \right]{\mathbf{v}}\left[ i \right],i \in {{\mathcal{I}}}$ or ${h_k}\left[ i \right] + {\mathbf{g}}_k^H\left[ i \right]{\mathbf{\Theta v}}\left[ i \right],i \in {{\mathcal{I}}}$. The former expression is applied when $T \approx \delta $, which means that the IRS can be reconfigured only once for each fading state. The latter expression is applied when $T \ll \delta $, which means that the IRS can be reconfigured only once for several fading states. How to extend our results in this paper to characterize the ergodic capacity over fading channels is an interesting problem worthy of further investigation.
  \item \emph{Capacity characterization with multiple APs and IRSs:} This paper considered the basic scenario with one IRS deployed for assisting the communication of one AP. Due to the limited coverage of IRS and the wide distribution of users, in practice, there may be a need to deploy multiple APs and IRSs. How to achieve globally optimal transmission is another interesting direction to be investigated in future work. Specifically, when there are multiple APs, users would suffer from interference caused by other unintended APs, which further complicates the optimization problem. In this case, other sophisticated mathematical tools are expected to be employed.
\end{itemize}
\vspace{-0.6cm}
\section*{Appendix~A: Proof of Theorem \ref{tsproof}} \label{Appendix:A}
\vspace{-0.3cm}
For the case of $N \to \infty$, the average achievable rate of user $k$ over the entire period $T$ can be rewritten as $\overline R _k^{\rm{N}} = \frac{1}{T}\int_0^T {R_k^{\rm{N}}\left( t \right)} dt$, where ${R_k^{\rm{N}}\left( t \right)}$ is obtained by replacing the discrete index $\left[ n \right]$ in \eqref{rate NOMA} with the continuous time index $\left( t \right)$. Then, problem (P1) for $N \to \infty $ can be equivalently rewritten as the following problem with continuous time variables
\vspace{-0.5cm}
\begin{subequations}\label{P1 ts}
\begin{align}
&\mathop {\max }\limits_{R^{\rm{N}},\left\{ {{\mathbf{\Theta}} \left( t \right),{p_k}\left( t \right)} \right\}} \;\;R^{\rm{N}}  \\
\label{rate profile ts}{\rm{s.t.}}\;\;&\frac{1}{T}\int_0^T {R_k^{\rm{N}}\left( t \right)} dt \ge {\alpha _k}R^{\rm{N}},\forall k,\\
\label{theta power ts}&{\mathbf{\Theta}}\left( t \right) \in \overline {\mathcal{S}},{p_k}\left( t \right)\in \overline {\mathcal{P}}, \forall t,k,
\end{align}
\end{subequations}
\vspace{-1.2cm}

\noindent where \eqref{rate profile ts} denotes the rate-profile constraint with continuous time variables. Here, for ease of exposition, $\overline {\mathcal{S}}$ denotes the feasible sets of $\left\{ {{\mathbf{\Theta }}\left( t \right)} \right\}$ specified by the discrete phase shift constraint \eqref{discrete phase shift NOMA} and the user decoding order constraint \eqref{SIC condition} for continuous time variables. Similarly, $\overline {\mathcal{P}}$ denotes the feasible sets of $\left\{ {{p_k}\left( t \right)} \right\}$ specified by the power allocation constraints \eqref{total power1 NOMA} and \eqref{total power2 NOMA} for continuous time variables. It is worth noting that $\overline {\mathcal{S}}$ and $\overline {\mathcal{P}}$ are not necessary convex set.\\
\indent Let $\left\{ {{\mathbf{\Theta }}_x^*\left( t \right),p_{k,x}^*\left( t \right)} \right\}$ and $\left\{ {{\mathbf{\Theta }}_y^*\left( t \right),p_{k,y}^*\left( t \right)} \right\}$ denote optimal solutions of problem \eqref{P1 ts} with optimal values $R_x^{*{\rm{N}}}$ and $R_y^{*{\rm{N}}}$, respectively. Therefore, $\left\{ {{\mathbf{\Theta }}_x^*\left( t \right),p_{k,x}^*\left( t \right)} \right\}$ and $\left\{ {{\mathbf{\Theta }}_y^*\left( t \right),p_{k,y}^*\left( t \right)} \right\}$ satisfy the following rate-profile constraints with $R_x^{*{\rm{N}}}$ and $R_y^{*{\rm{N}}}$
\vspace{-0.4cm}
\begin{align}\label{rp xy}
\frac{1}{T}\int_0^T {R_{k,i}^{*\rm{N}}\left( t \right)} dt \ge {\alpha _k}R_i^{*\rm{N}},\forall k,
\end{align}
\vspace{-1.2cm}

\noindent where ${R_{k,i}^{*\rm{N}}\left( t \right)}$ denotes the corresponding optimal instantaneous communication rate and $i\! \in\! \left\{ {x,y} \right\}$.\\
\indent To show that problem \eqref{P1 ts} satisfies the time-sharing condition in \cite{timeshare}, for any $0 \le \nu  \le 1$, we need to construct feasible solutions $\left\{ {{{\mathbf{\Theta }}_z}\left( t \right),{p_{k,z}}\left( t \right)} \right\}$ such that they (i) satisfy the rate-profile constraints with $\nu R_x^{*N} + \left( {1 - \nu } \right)R_y^{*N}$; and (ii) achieve an average sum rate equal or higher than $\nu R_x^{*N} + \left( {1 - \nu } \right)R_y^{*N}$. Note that, for problem \eqref{P1 ts}, the two conditions (i) and (ii) are equivalent, i.e., if one condition is met, the other is automatically satisfied as well. Such $\left\{ {{{\mathbf{\Theta }}_z}\left( t \right),{p_{k,z}}\left( t \right)} \right\}$ can be constructed by allocating $\nu $ percentage of the entire period $T$ for solutions $\left\{ {{\mathbf{\Theta }}_x^*\left( t \right),p_{k,x}^*\left( t \right)} \right\}$ and $\left( {1 - \nu } \right)$ percentage of the entire period $T$ for solutions $\left\{ {{\mathbf{\Theta }}_y^*\left( t \right),p_{k,y}^*\left( t \right)} \right\}$ as follows
\vspace{-0.3cm}
\begin{align}
{{\mathbf{\Theta }}_z}\left( t \right) = \left\{ \begin{gathered}
  {\mathbf{\Theta }}_x^*\left( {\frac{t}{\nu }} \right),0 \le t \le \nu T \hfill \\
  {\mathbf{\Theta }}_y^*\left( {\frac{{t - \nu T}}{{1 - \nu }}} \right),\nu T < t \le T \hfill \\
\end{gathered}  \right.,{p_{k,z}}\left( t \right) = \left\{ \begin{gathered}
  p_{k,x}^*\left( {\frac{t}{\nu }} \right),0 \le t \le \nu T \hfill \\
  p_{k,y}^*\left( {\frac{{t - \nu T}}{{1 - \nu }}} \right),\nu T < t \le T \hfill \\
\end{gathered}  \right.,
\end{align}
\vspace{-0.7cm}

\noindent Accordingly, the instantaneous communication rate of user $k$ achieved by the above constructed solutions $\left\{ {{{\mathbf{\Theta }}_z}\left( t \right),{p_{k,z}}\left( t \right)} \right\}$ is given by
\vspace{-0.3cm}
\begin{align}\label{rate z}
R_{k,z}^N\left( t \right) = \left\{ \begin{gathered}
  R_{k,x}^{*N}\left( {\frac{t}{\nu }} \right),0 \le t \le \nu T \hfill \\
  R_{k,y}^{*N}\left( {\frac{{t - \nu T}}{{1 - \nu }}} \right),\nu T < t \le T \hfill \\
\end{gathered}  \right.,\forall k.
\end{align}
\vspace{-0.7cm}

\noindent Then, the corresponding rate-profile constraints can be expressed as
\vspace{-0.3cm}
\begin{align}\label{rate-profile z}
\begin{gathered}
  \frac{1}{T}\int_0^T {R_{k,z}^N\left( t \right)} dt = \frac{1}{T}\left( {\int_0^{\nu T} {R_{k,x}^{*N}\left( {\frac{t}{\nu }} \right)} dt + \int_{\nu T}^T {R_{k,y}^{*N}\left( {\frac{{t - \nu T}}{{1 - \nu }}} \right)} dt} \right) \hfill \\
  \mathop  = \limits^{\left( a \right)} \frac{\nu }{T}\int_0^T {R_{k,x}^{*N}\left( \omega  \right)} d\omega  + \frac{{1 - \nu }}{T}\int_0^T {R_{k,y}^{*N}\left( \tau  \right)} d\tau  \mathop  \ge \limits^{\left( b \right)} {\alpha _k}\left( {\nu R_x^{*N} + \left( {1 - \nu } \right)R_y^{*N}} \right),\forall k, \hfill \\
\end{gathered}
\end{align}
\vspace{-0.7cm}

\noindent where (a) is obtained by replacing $t$ in the first and second term with $t = \nu \omega $ and $t = \left( {1 - \nu } \right)\tau  + \nu T$, respectively, and (b) holds due to Equation \eqref{rp xy}. Equation \eqref{rate-profile z} means that the constructed solutions satisfy conditions (i) and (ii). Therefore, problem \eqref{P1 ts} satisfies the time-sharing condition in \cite{timeshare}, which implies that the maximum value of the optimization problem \eqref{P1 ts} is a concave function of $R^{\rm{N}}$ even though $\overline {\mathcal{S}}$ and $\overline {\mathcal{P}}$ are all non-convex. It is also worth mentioning that the construction of $\left\{ {{{\mathbf{\Theta }}_z}\left( t \right),{p_{k,z}}\left( t \right)} \right\}$ for satisfying the time-sharing condition is only valid with continuous time variables (i.e., $N \to \infty$), which in general does not hold for the case of finite $N$. Thus, the proof of Theorem \ref{tsproof} is completed.
\vspace{-0.6cm}
\section*{Appendix~B: Proof of Proposition \ref{two user NOMA}} \label{Appendix:B}
\vspace{-0.4cm}
For the three user case, we first consider the scenario where there is only one active user and the maximum of ${\phi ^{\left( {\left\{ {\lambda _k^{{\rm{N}}}} \right\},{\mathbf{\Theta}} } \right)}}\left( {\left\{ {{q_k}} \right\}} \right)$ is achieved on the vertexes of ${\left( {{P_{\max}},0,0} \right)_{\left\{ 1 \right\}}},{\left( {{P_{\max}},{P_{\max}},0} \right)_{\left\{ 2 \right\}}}$,\\${\left( {{P_{\max}},{P_{\max}},{P_{\max}}} \right)_{\left\{ 3 \right\}}}$, where the subscript represents the active user index.\\
\indent Next, when there are two active users, the constraint on power allocation becomes ${P_{\max }} = {q_j} > {q_k} > 0,\forall j < k \in {\mathcal{K}}$. Now, the maximum of ${\phi ^{\left( {\left\{ {\lambda _k^{{\rm{N}}}} \right\},{\mathbf{\Theta}} } \right)}}\left( {\left\{ {{q_k}} \right\}} \right)$ is achieved at the stationary point ${\overline q _k}$. Then, we obtain \eqref{q0} by solving ${\nabla _{{q_k}}}{\phi ^{\left( {\left\{ {\lambda _k^{{\rm{N}}}} \right\},{\mathbf{\Theta}} } \right)}}\left( {\left\{ {{q_k}} \right\}} \right) = 0$. The stationary points for two active users are ${\left( {{P_{\max }},{P_{\max }},{{\overline q }_3}} \right)_{\left\{ {2,3} \right\}}},{\left( {{P_{\max }},{{\overline q }_2},0} \right)_{\left\{ {1,2} \right\}}},{\left( {{P_{\max }},{{\overline q }_3},{{\overline q }_3}} \right)_{\left\{ {1,3} \right\}}}$.\\
\indent Then, for the general three active users case, the constraint on power allocation becomes ${P_{\max }} = {q_1} > {q_2} > {q_2} > 0$. The corresponding stationary point is ${\left( {{P_{\max }},{{\overline q }_2},{{\overline q }_3}} \right)_{\left\{ {1,2,3} \right\}}}$.\\
\indent Hence, the proof of Proposition \ref{two user NOMA} with three users is completed. The proof for two users is similar and we omit it for brevity.
\vspace{-0.6cm}
\section*{Appendix~C: Proof of Lemma \ref{D OMA 1}} \label{Appendix:C}
\vspace{-0.4cm}
Under a given IRS reflection matrix ${\mathbf{\Theta}}$, we can rewrite each subproblem in \eqref{D OMA} as follows
\vspace{-0.3cm}
\begin{subequations}\label{D OMA Appendix}
\begin{align}\label{D OMA Appendix obj}
\mathop {\max }\limits_{\left\{ {{p_k} ,{\omega _k} } \right\}} &\;\sum\nolimits_{k = 1}^K {\frac{{\lambda _k^{{\rm{O}}}}}{T}} {\omega _k}{\log _2}\left( {1 + \frac{{{{\left| {{h_k} + {\mathbf{g}}_k^H{\mathbf{\Theta}} {\mathbf{v}}} \right|}^2}{p_k}}}{{{\omega _k}{\sigma ^2}}}} \right)  \\
\label{D OMA Appendix1}{\rm{s.t.}}\;\;&\sum\nolimits_{k = 1}^K {{p_k}}  \le {P_{\max }},\\
\label{D OMA Appendix3}&\sum\nolimits_{k = 1}^K {{\omega _k}} \le 1,\\
\label{D OMA Appendix4}&{p_k} \ge 0, 0 \le {\omega _k} \le 1,\forall k,
\end{align}
\end{subequations}
\vspace{-1cm}

\noindent Define ${\omega _k}{\log _2}\left( {1 + \frac{{{{\left| {{h_k} + {\mathbf{g}}_k^H{\mathbf{\Theta}} {\mathbf{v}}} \right|}^2}{p_k}}}{{{\omega _k}{\sigma ^2}}}} \right) \triangleq 0$ when ${\omega _k} = 0,\forall k$, such that the objective function of \eqref{D OMA Appendix obj} is jointly concave with respect to ${\omega _k}$ and ${p_k}$. Therefore, problem \eqref{D OMA Appendix} is a convex problem and we apply the Lagrangian dual method to optimally solve it. New non-negative Lagrange multipliers ${\delta ^{{\rm{O}}}}$ and ${\nu ^{{\rm{O}}}}$ are introduced associated with constraints \eqref{D OMA Appendix1} and \eqref{D OMA Appendix3}, respectively. For given ${\delta ^{{\rm{O}}}}$ and ${\nu ^{{\rm{O}}}}$, the Lagrange dual function of problem \eqref{D OMA Appendix} can be expressed as
\vspace{-0.4cm}
\begin{subequations}\label{Lagrange dual function D OMA Appendix}
\begin{align}
\mathop {\max }\limits_{\left\{ {{p_k}} \right\},\left\{ {{\omega _k}} \right\}} &{{{\mathcal{L}}}_3}\left( {\left\{ {{p_k}} \right\},\left\{ {{\omega _k}} \right\},{\delta ^{{\rm{O}}}},{\nu ^{{\rm{O}}}}} \right) \\
\label{Lagrange dual function D OMA Appendix1}{\rm{s.t.}}\;\;&\eqref{D OMA Appendix4},
\end{align}
\end{subequations}
\vspace{-1.2cm}

\noindent where ${{{\mathcal{L}}}_3}\left(\! {\left\{ {{p_k}} \right\},\!\left\{ {{\omega _k}} \right\},\!{\delta ^{{\rm{O}}}},\!{\nu ^{{\rm{O}}}}} \!\right) \!= \!\sum\nolimits_{k = 1}^K {\frac{{\lambda _k^{{\rm{O}}}}}{T}} {\omega _k}{\log _2}\left( {1 + \frac{{{{\left| {{h_k}\! + \!{\mathbf{g}}_k^H{\mathbf{\Theta}} {\mathbf{v}}} \right|}^2}{p_k}}}{{{\omega _k}{\sigma ^2}}}} \right) - {\delta ^{{\rm{O}}}}\sum\nolimits_{k = 1}^K {{p_k}}  - {\nu ^{{\rm{O}}}}\sum\nolimits_{k = 1}^K {{\omega _k}} $.\\
\indent Note that the problem \eqref{Lagrange dual function D OMA Appendix} is jointly concave with respect to ${\omega _k}$ and ${p_k}$, hence, the Karush-Kuhn-Tucker (KKT) conditions are necessary and sufficient for the optimality of \eqref{Lagrange dual function D OMA Appendix}. By taking the derivative of the objective function of \eqref{Lagrange dual function D OMA Appendix} with respect to ${p_k^*}$, the optimal power allocation structure to \eqref{Lagrange dual function D OMA Appendix} under given ${\delta ^{{\rm{O}}}}$ and ${\nu ^{{\rm{O}}}}$ proves to be $p_k^* = \omega _k^*t_k^{{\rm{O}}},\forall k$, where $t_k^{{\rm{O}}} = {\left( {\frac{{\lambda _k^{{\rm{O}}}}}{{{\delta ^{{\rm{O}}}}T\ln 2}} - \frac{{{\sigma ^2}}}{{{{\left| {{h_k} + {\mathbf{g}}_k^H{\mathbf{\Theta}} {\mathbf{v}}} \right|}^2}}}} \right)^ + }$. Though the optimal values of ${p_k^*}$ and ${\omega _k^*}$ are coupled, the value of $t_k^{{\rm{O}}}$ is uniquely determined by the dual variables. By substituting ${p_k^*}$ into \eqref{D OMA Appendix}, we get
\vspace{-0.4cm}
\begin{subequations}\label{D OMA Appendix equal}
\begin{align}
\mathop {\max }\limits_{\left\{ {{\omega _k}} \right\}} &\;\sum\nolimits_{k = 1}^K {{\omega _k}{g_k}\left( {t_k^{{\rm{O}}}} \right)}   \\
\label{D OMA Appendix equal1}{\rm{s.t.}}\;\;&\sum\nolimits_{k = 1}^K {{\omega _k}} \le 1, 0 \le {\omega _k} \le 1,\forall k,
\end{align}
\end{subequations}
\vspace{-0.8cm}

\noindent where ${g_k}\left( {t_k^{{\rm{O}}}} \right) = \frac{{\lambda _k^{{\rm{O}}}}}{T}{\log _2}\left( {1 + \frac{{{{\left| {{h_k} + {\mathbf{g}}_k^H{\mathbf{\Theta}} {\mathbf{v}}} \right|}^2}t_k^{{\rm{O}}}}}{{{\sigma ^2}}}} \right)$. It is evident that problem \eqref{D OMA Appendix equal} is a LP whose optimal solutions are given by
\vspace{-0.3cm}
\[p_k^* = \left\{ \begin{gathered}
  {\left( {\frac{{\lambda _k^{{\rm{O}}}}}{{{\delta ^{{\rm{O}}}}T\ln 2}} - \frac{{{\sigma ^2}}}{{{{\left| {{h_k} + {\mathbf{g}}_k^H{\mathbf{\Theta}} {\mathbf{v}}} \right|}^2}}}} \right)^ + },{\rm{if}}\;\;k = {k^*} \hfill \\
  0,\;{\rm {otherwise}} \hfill \\
\end{gathered}  \right.,\omega _k^* = \left\{ \begin{gathered}
  1,{\rm{if}}\;\;k = {k^*} \hfill \\
  0,\;{\rm {otherwise}} \hfill \\
\end{gathered}  \right.\]
\vspace{-0.7cm}

\noindent where ${k^*} = \arg \mathop {\max }\limits_{k \in {{\mathcal{K}}}} \;{g_k}\left( {t_k^{{\rm{O}}}} \right)$, which indicates that there is only one user served at the optimal solution. By updating ${{\delta ^{{\rm{O}}}}}$ until $p_k^* = {P_{\max }}$, the optimal $\left\{ {p_k^*,\omega _k^*} \right\}$ to problem \eqref{D OMA Appendix} under given ${\mathbf{\Theta}} $ is achieved among $\left\{ {\left( {{{\mathbf{0}}_{k - 1}},{P_{\max }},{{\mathbf{0}}_{K - k}}} \right),\left( {{{\mathbf{0}}_{k - 1}},1,{{\mathbf{0}}_{K - k}}} \right),\forall k} \right\}$ leading to a larger objective value. It is evident that the optimal IRS reflection matrix for the $k$th solution should satisfy ${{\mathbf{\Theta}} _k} = \arg \mathop {\max }\limits_{{\mathbf{\Theta}}  \in {\mathcal{S}}} \;\;{\left| {{h_k} + {\mathbf{g}}_k^H{\mathbf{\Theta}} {\mathbf{v}}} \right|^2},\forall k$. Hence, we complete the proof for Lemma \ref{D OMA 1}.

\vspace{-0.4cm}
\bibliographystyle{IEEEtran}
\bibliography{mybib}

\end{document}